\documentstyle[12pt,a41,epsfig,cite,amssymb,amstex]{article}
\newcommand{\MS}{\overline{{\sf MS}}}
\newcommand{\Ahathat}{\hat{\hspace*{-1mm}\hat{A}}}
\newcommand{\bra}[1]{\langle#1\mid}
 \newcommand{\ket}[1]{\mid#1\rangle}

 \newcommand{\N}{\nonumber}

\newcommand{\bea}{\begin{eqnarray}}
\newcommand{\bq}{\begin{equation}}
\newcommand{\eea}{\end{eqnarray}}
\newcommand{\eq}{\end{equation}}

\newcommand\be{\begin{eqnarray}}
\newcommand\ee{\end{eqnarray}}

\newcommand\ep{\varepsilon}

\newcommand\Li{{\rm Li}}

\begin{document}
\noindent
\sloppy
\thispagestyle{empty}

\begin{flushleft}
DESY 09--60
\hfill {\tt arXiv:0909.1547}
\\
SFB/CPP--09--073\\
September 2009
\end{flushleft}

\vspace*{\fill}
\begin{center}
{\LARGE\bf  \boldmath{$O(\alpha_s^2)$}  and \boldmath{$O(\alpha_s^3$)}  Heavy 
Flavor Contributions} 

\vspace{2mm}
{\LARGE\bf to  Transversity at \boldmath{$Q^2 \gg m^2$}}

\vspace{2cm}
\large
Johannes Bl\"umlein,  Sebastian Klein, and Beat T\"odtli
\\
\vspace{2em}
\normalsize
{\it Deutsches Elektronen--Synchrotron, DESY,\\
Platanenallee 6, D--15738 Zeuthen, Germany}
\\
\vspace{2em}
\end{center}
\vspace*{\fill}
%
\begin{abstract}
\noindent
In deep-inelastic processes the heavy flavor Wilson coefficients factorize for 
$Q^2 \gg m^2$ into the light flavor Wilson coefficients of the corresponding 
process and the massive operator matrix elements (OMEs). We calculate the 
$O(\alpha_s^2)$ and $O(\alpha_s^3)$ massive OME for the flavor non-singlet 
transversity distribution. At $O(\alpha_s^2)$ the OME is obtained for general 
values of the Mellin variable $N$, while at $O(\alpha_s^3)$ the moments $N = 1$ to 
13 are computed. The terms $\propto T_F$ of the 3--loop transversity anomalous 
dimension are obtained and results in the literature are confirmed. We discuss the 
relation of these contributions to the Soffer bound for transversity.
\end{abstract}
\vspace*{\fill}
\newpage
\section{Introduction}
\label{sec-1}

\vspace{1mm}\noindent
The transversity distribution $\Delta_T f(x,Q^2)$ is one of the three 
possible quarkonic twist-2 parton distributions besides the unpolarized 
quark density $f(x,Q^2)$ and the longitudinally polarized  density 
$\Delta f(x,Q^2)$. Unlike the latter distributions, it cannot be measured in 
inclusive deeply inelastic scattering since the corresponding contribution 
is $\propto m_q^2/Q^2$, \cite{FEYN}, with $m_q$ a light quark mass and $Q^2$ 
the virtuality of the exchanged gauge boson. It can be extracted from 
deep-inelastic scattering studying isolated meson production, also called 
semi-inclusive deeply-inelastic scattering (SIDIS), \cite{PROC1,PROC2}, 
and in the polarized Drell-Yan process~\cite{PROC2,PROC3,ARTR}.~\footnote{For 
a review see Ref.~\cite{RATCL}.} Measurements of the transversity distribution in 
different polarized hard scattering processes are currently performed or in 
preparation \cite{EXP}. In the past phenomenological models for the transversity 
distribution were developed based on bag-like models, chiral models, light-cone 
models, spectator models, and non-perturbative QCD calculations, cf. \cite{PHEN}. 
The main behaviour of the distributions is that they vanish by some power law 
both at small and large values of Bjorken $x$ and exhibit a shifted bell-like 
shape. First attempts to extract the distributions out of data were made in 
Refs.~\cite{ANSELM}.
The moments of the transversity distribution can be measured in lattice 
simulations, which help to constrain it ab initio. First results were given in 
Refs.~\cite{LATT}. From these investigations there is evidence, that the 
up-quark
distribution is positive while the down-quark distribution is negative, with first
moments between 0.85 \ldots 1.0 and --0.20 \ldots --0.24, respectively. 

The scaling violations of the transversity distribution were explored
in leading-, \cite{ARTR,LO,LO1,LO2}~\footnote{The small $x$ limit of the 
LO anomalous dimension was calculated in \cite{LXLO}.}, and next-to-leading order, 
\cite{NLO,NLO1,NLO2}.~\footnote{For calculations in the non-forward case see 
\cite{NFW,LO1}.} In Ref.~\cite{LO1} also the method proposed in \cite{IK}
was used to calculate the anomalous dimension. At three-loop order the moments 
$N = 1$ to 8 for the anomalous dimension are known 
\cite{GRAC}. For the calculation
of the scattering cross sections also the corresponding Wilson coefficients 
have to be known. In case of SIDIS these corrections have not yet been 
calculated. For the transversely polarized Drell-Yan process the $O(\alpha_s)$ 
Wilson 
coefficient
was derived in Ref.~\cite{NLO1} based on \cite{VW} and at higher orders 
the contributions due to soft-resummation are available \cite{soft}. 

The scattering cross sections dominated by the transversity distribution receive 
heavy flavor corrections, although transversity itself is a flavor non-singlet 
distribution. These contributions reside in the corresponding Wilson coefficients. 
In deep-inelastic processes the heavy flavor Wilson coefficients factorize into 
massive operator matrix elements (OMEs) and the light flavor Wilson coefficients 
at large enough momentum transfer $Q^2 \gg m^2$, as was shown in 
Ref.~\cite{BUZA1}, with $m$ the heavy quark mass. 
In this way all contributions except the power corrections $(m^2/Q^2)^k,~~k \geq 1$
can be calculated. The massive OMEs derive from the twist 2 operators emerging 
in the light--cone expansion between on--shell states and are process 
independent quantities. The formalism proposed in Ref.~\cite{BUZA1}
has been applied successfully to calculate 
the asymptotic heavy flavor 
Wilson coefficients at $O(\alpha_s^2)$ \cite{BUZA1,AHF2,BBK1}
in unpolarized and polarized deep-inelastic scattering. For $F_L(x,Q^2)$ the 
asymptotic heavy flavor corrections to $O(\alpha_s^3)$ were calculated in 
Ref.~\cite{BFNK}. A series of Mellin moments for the asymptotic heavy flavor 
Wilson 
coefficients contributing to the structure function $F_2(x,Q^2)$ at 
$O(\alpha_s^3)$ have recently been computed in 
Refs.~\cite{BBK3,AHF3}.

In the present paper we apply this formalism to the tensor-operator defining the 
flavor non-singlet transversity distribution, and limit the consideration to 
contributions of 
twist 2 and 
the collinear parton model. We calculate the $O(\alpha_s^2)$ corrections for 
the 
flavor non-singlet OME of transversity
for general values of the Mellin variable $N$. At $O(\alpha_s^3)$ the OME is 
computed for 
individual Mellin moments $N = 1$ to $13$. The 2-loop calculation verifies the 
$T_F$-terms of the transversity anomalous dimension of former NLO calculations 
\cite{NLO,NLO1,NLO2}. In the 3-loop 
calculation 
we obtain the moments for the complete 2-loop anomalous dimension, 
which appears in the
double pole term in the dimensional parameter $\varepsilon = D - 4$. Furthermore, the 
$T_F$-contributions to the 3-loop anomalous dimension are obtained from the 
single pole term, which can be compared to the results in \cite{GRAC} for 
$N=1$ to  
$8$, while  the $T_F$-terms of the anomalous dimension for $N = 9$ to $13$ are new.
The results for the massive OME for transversity given in the 
present paper are related to future lattice simulations 
with (2+1+1)-, resp. (2+1)-dynamical fermions. The heavy flavor contributions 
are also of 
relevance for the Soffer bound for transversity \cite{SOFFER}. 

The paper is organized as follows. In Section~2 we summarize the main 
relations for semi--inclusive scattering cross sections in the leading 
twist approximation
from which the transversity distribution can be determined. 
Here, as in the case of inclusive deep-inelastic scattering, tagging on 
charm-mesons 
allows to measure the charm contribution directly in high-luminosity 
experiments. The method to calculate the heavy flavor corrections in the 
asymptotic region is briefly described. In Section~3 we
calculate the $O(\alpha_s^2)$ massive operator matrix element. The Mellin 
moments
of the OME at $O(\alpha_s^3)$ are computed in Section~4. In Section~5 we 
discuss the
heavy flavor contributions to the Soffer bound and Section~6 contains the 
conclusions.
In the Appendix we summarize the $T_F$-parts of the 3--loop anomalous 
dimension for 
transversity and the moments  of the constant part $O(\varepsilon^0)$ of the 
un-renormalized $O(\alpha_s^3)$ massive OME for the Mellin moments $N =1$ to 
$13$. For details concerning the calculation and renormalization of
massive non-singlet OMEs we refer to Ref.~\cite{BBK3}.
\section{\boldmath Basic Formalism}
\label{sec-2}

\vspace{1mm}\noindent
The transversity distribution
\begin{eqnarray}
\label{eqTR1}
\Delta_T f(x,Q^2) \equiv f^{\uparrow}(x,Q^2) - f^{\downarrow}(x,Q^2)
\end{eqnarray}
contributes to a large variety of scattering processes, cf.~\cite{RATCL}.
Here $\uparrow (\downarrow)$ denote the transverse spin directions. Eq.~(\ref{eqTR1})
describes the transversity distribution obtained in the light--cone expansion
at twist 2 or in the collinear parton model. For other phenomenological 
applications 
one may introduce $k_\perp$--effects for this distribution, \cite{RATCL}. This, however, 
has consequences for the twist expansion and the renormalization of the corresponding
processes, when calculating them to higher orders. We will therefore restrict 
the 
analysis to the level of twist 2 and consider only processes  which are free
of $k_\perp$--effects, or after these were integrated out in the final state.

For semi-inclusive deeply inelastic charged lepton-nucleon scattering 
$l N \rightarrow l' h + X$
the Born cross section, after the ${\bf P}_{h \perp}$-integration,
is given by, \cite{RATCL},
\begin{eqnarray}
\label{sidis1}
\frac{d^3 \sigma}{dx dy dz} &=& \frac{4 \pi \alpha_{\rm em}^2 s}{Q^4} 
\sum_{a =q,\overline{q}} e_a^2 x \Biggl\{\frac{1}{2} \left[1+(1-y)^2\right]
F_a(x,Q^2) \tilde{D}_a(z,Q^2) \nonumber\\ &&
- (1-y) |{\bf S}_\perp||{\bf S}_{h\perp}| \cos\left(\phi_S + \phi_{S_h}\right)
\Delta_T F_a(x,Q^2) \Delta_T \tilde{D}_a(z,Q^2)\Biggr\}~.
\end{eqnarray}
   Here, in addition to the Bjorken variables $x$ and $y$, the
   fragmentation variable $z$ occurs.
    ${\bf S}_\perp$ and ${\bf S}_{h\perp}$ are the 
   transverse spin vectors of the incoming nucleon $N$ and the measured hadron 
   $h$. The angles $\phi_{S,S_h}$ are measured in the plane perpendicular to
   the $\gamma^* N$ ($z$--) axis between the $x$--axis and the respective 
vector.
   The transversity distribution can be obtained from Eq.~(\ref{sidis1}) for a 
   {\sf transversely} polarized hadron $h$ by measuring its polarization. 
   The functions $F_i, \tilde{D}_i, \Delta_T F_i, \Delta_T \tilde{D}_i$ are
   given by
   \begin{eqnarray}
\label{eqstr1}
    F_i(x,Q^2) &=& {\cal C}_i(x,Q^2) \otimes f_i(x,Q^2)~, \\
    \tilde{D}_i(z,Q^2)  &=& \tilde{{\cal C}}_i(z,Q^2) \otimes {D}_i(z,Q^2)~, 
\\
\label{eqstr2}
    \Delta_T F_i(x,Q^2) &=& \Delta_T {\cal C}_i(x,Q^2) \otimes 
                            \Delta_T f_i(x,Q^2)~,\\
    \Delta_T \tilde{D}_i(z,Q^2)  &=& \Delta_T \tilde{{\cal C}}_i(z,Q^2) 
                                     \otimes \Delta_T {D}_i(z,Q^2)~.
   \end{eqnarray}
   Here, $\otimes$ denotes the Mellin convolution,
$D_i, \Delta_T {D}_i$ are the fragmentation functions and 
   ${{\cal C}}_i,~\tilde{{\cal C}}_i,~\Delta_T {\cal C}_i,
   ~\Delta_T \tilde{{\cal C}}_i$ 
   are the corresponding
   space- and time-like Wilson coefficients. 
   The
   Wilson coefficient for transversity, $\Delta_T {\cal C}_i(x,Q^2)$,
   contains light  ($\Delta_T C_i$)
and heavy flavor ($\Delta_T H_i$) contributions
   \begin{eqnarray}
\label{eqWIL1} 
   \Delta_T {\cal C}_i(x,Q^2) = \Delta_T C_i(x,Q^2) 
                    + \Delta_T H_i(x,Q^2)~.
   \end{eqnarray}
For brevity we dropped arguments like $m^2$, the factorization scale, 
$\mu^2$, and the number of light flavors, $N_f$, in Eq.~(\ref{eqWIL1}).

   Eq.~(\ref{sidis1}) holds for spin--$1/2$ hadrons in the final state, but
   the transversity distribution may also be measured in the lepto-production
   process of spin--1 hadrons, \cite{JI}. In this case, the 
   ${\bf P}_{h \perp}$-integrated Born cross section reads 
   \begin{eqnarray}
     \frac{d^3\sigma}{dxdydz}&=&\frac{4\pi\alpha^2}{xyQ^2} 
         \sin\left(\phi_S + \phi_{S_{LT}}\right)
             |{\bf S}_\perp||{S}_{LT}| (1-y)
            \sum_{i =q,\overline{q}} e_i^2 
                x \Delta_T F_i(x,Q^2) \widehat{H}_{i,1,LT}(z,Q^2)~.
    \label{sidis2}
   \end{eqnarray}
   Here, the polarization state of a spin--1 particle is described by a tensor 
   with five independent components, \cite{BAMU}. $\phi_{LT}$ 
   denotes the azimuthal angle of $\vec{S}_{LT}$, with 
   \begin{eqnarray}
    |S_{LT}| = \sqrt{\left(S_{LT}^x\right)^2 +\left(S_{LT}^y\right)^2}~. 
   \end{eqnarray}
   $\widehat{H}_{a,1,LT}(z,Q^2)$ is a $T$- and chirally odd twist-2 
   fragmentation function at vanishing $k_\perp$. Process (\ref{sidis2}) 
   has the advantage that the transverse polarization of the produced hadron 
   can be measured from its decay products. 

   The transversity distribution can also be measured in the transversely 
   polarized Drell--Yan process using the polarization asymmetry, see 
   Refs.~\cite{VW,NLO1,soft}.  However, the 
   SIDIS processes have the advantage that in 
   high luminosity experiments, cf. \cite{EIC}, the heavy flavor contributions 
can be tagged 
   like in deep-inelastic scattering. This is 
   not the case for the Drell-Yan process, where the heavy flavor effects
   appear as inclusive radiative corrections in the Wilson coefficients.
   We will therefore mainly consider SIDIS in the following.

   As was shown in Ref.~\cite{BUZA1}, in the region $Q^2 \gg 
m^2$ all non--power contributions to the heavy quark Wilson coefficients
obey factorization relations. 
In the general flavor non-singlet case one obtains for $N_f$ light  and 
one heavy quark 
\begin{eqnarray}
\label{HFAC}
H_{a}^{\rm asymp, NS} \left(x,\frac{Q^2}{m^2}, \frac{m^2}{\mu^2}\right) = 
C_{a,q}^{\rm NS} \left(x,\frac{Q^2}{\mu^2}, N_f+1\right) 
\otimes	                           A_{qq,Q}^{\rm NS} 
\left(x,\frac{m^2}{\mu^2}\right) 
-C_{a,q}^{\rm NS} \left(x,\frac{Q^2}{\mu^2}, N_f\right)~, 
\end{eqnarray}
where $C_{a,q}^{\rm NS}$ is a
light flavor Wilson coefficient
 and $A_{qq,Q}^{\rm NS}$ is the corresponding 
massive operator matrix element, cf.~\cite{BUZA1,BBK1,BBK3},
with
\begin{eqnarray}
\label{coeff}
C_{a,q}^{\rm NS} \left(x,\frac{Q^2}{\mu^2}\right) &=& \delta(1-x) +
\sum_{k=1}^\infty a_s^k(\mu^2)
C_{a,q}^{(k),\rm NS} \left(x, \frac{Q^2}{\mu^2}\right)~, \\
\label{ome}
A_{qq,Q}^{\rm NS} \left(x,\frac{m^2}{\mu^2}\right) &=& 
\langle q| O^{\rm NS}|q \rangle =
\delta(1-x) +\sum_{k=2}^\infty a_s^k(\mu^2) A_{qq,Q}^{(k),\rm NS} 
\left(x,\frac{m^2}{\mu^2}\right)~.
\end{eqnarray}
Here $a_s(\mu^2) = \alpha_s(\mu^2)/(4\pi)$ denotes the strong coupling 
constant and
$|q\rangle$ are light quark states, with on-shell momenta.
The local flavor non-singlet twist-2 operator for transversity is given by 
\begin{eqnarray}
\label{op2}
O_{q,r}^{{\rm TR,NS}, \mu, \mu_1, \ldots, \mu_N}(z) &=& \frac{1}{2} i^{N-1}  
S \left[\overline{q}(z) \sigma^{\mu \mu_1}
D^{\mu_2} \ldots D^{\mu_N} \frac{\lambda_r}{2}
q(z)\right] - {\sf Trace~Terms}~,
\end{eqnarray}
with $\sigma^{\mu\nu} = (i/2)\left[\gamma^\mu \gamma^\nu -
\gamma^\nu \gamma^\mu \right]$, $\lambda_r$ the Gell-Mann matrices for 
$SU(3)_{\rm flavor}$, $D^{\nu}$ the covariant derivative in QCD,
$q (\overline{q})$ denote the quark and antiquark fields, and the operator $S$  
symmetrizes the Lorentz indices.
Note that in Eq.~(\ref{HFAC})
the heavy quark degrees of freedom are all contained in the process 
independent OMEs. 

In case of transversity one obtains the following 
representation for the heavy flavor Wilson coefficient~\footnote{Apparently, the light 
flavor Wilson coefficients for SIDIS were not yet calculated even at $O(a_s)$, 
although this calculation and the corresponding soft-exponentiation should be 
straightforward.} after
expanding Eq.~(\ref{HFAC}) up to  $O(a_s^3)$
   \begin{eqnarray}
     \Delta_T H^{{\rm asym}}_q(N_f+1)=&&
                 a_s^2(N_f+1)\Bigl[ \Delta_T A_{qq,Q}^{(2),\rm NS} 
                             +\Delta_T \hat{C}_q^{(2)}(N_f)
                      \Bigr]
\N \\
                &+&a_s^3(N_f+1)\Bigl[ 
                              \Delta_T A_{qq,Q}^{(3),\rm NS}(N_f+1)
                             +\Delta_T A_{qq,Q}^{(2),\rm NS}
                              \Delta_T C_q^{(1)}
\N\\ && \hspace{5mm}
                             +\Delta_T\hat{C}_q^{(3)}(N_f)\Bigr]~.
    \label{HwcofTR}
   \end{eqnarray}
Here we made the $N_f$-dependence explicit and  use the notation 
\begin{eqnarray}
\hat{f}(N_f) = f(N_f+1) - f(N_f)~.
\end{eqnarray}
We dropped all arguments like $x, N, m^2, \mu^2$, which are understood 
implicitly. Additionally, Eq.~(\ref{HwcofTR}) is written in Mellin space,
in which we will work from now on, if not stated otherwise. The assignment
of the differing arguments in $N_f$ in Eq.~(\ref{HwcofTR}) is necessary to 
project onto the heavy quark part.

Following Ref.~\cite{BBK3} we consider the Green's function $\hat{G}^{ij,{\rm 
TR,NS}}_{\mu,q,Q}$
which is obtained by contracting the matrix element of the local operator 
(\ref{op2})
with the source term $J_N = \Delta^{\mu_1} \ldots \Delta^{\mu_N}$ 
  \begin{eqnarray}
  \overline{u}(p,s) G^{ij, {\rm TR,NS}}_{\mu,q,Q} \lambda_r u(p,s)&=&
   J_N\bra{q_i(p)}O_{q,r;\mu, \mu_1, \ldots, \mu_N}^{{\rm TR,NS}}
        \ket{q^j(p)}_Q~\label{GijTRNS}~, 
  \end{eqnarray}
where $p$ and $s$ denote the 4--vectors of the momentum and spin 
of the external light quark line, $u(p,s)$ is the corresponding 
bi--spinor, $\Delta.\Delta = 0$, and $Q$ labels the heavy quark 
contribution.
The un-renormalized Green's function has the following Lorentz structure
  \begin{eqnarray}
   \hat{G}^{ij, {\rm TR,NS}}_{\mu,q,Q}&=&
            \delta_{ij}(\Delta \cdot p)^{N-1} 
             \Bigl(
                   \Delta_{\rho}\sigma^{\mu\rho}
                   \Delta_T~\Ahathat_{qq,Q}^{\rm NS}
                        \Bigl(\frac{\hat{m}^2}{\mu^2},\ep,N\Bigr) 
                  +c_1 \Delta^\mu + c_2 p^\mu 
                  +c_3 \gamma^\mu p \hspace*{-2mm} / 
\N\\ && \hspace{30mm}
                  +c_4 \Delta \hspace*{-3mm}/~p \hspace*{-2mm}/ \Delta^\mu
                  +c_5 \Delta \hspace*{-3mm}/~p \hspace*{-2mm}/ p^\mu
             \Bigr)~,
  \end{eqnarray}
with unphysical constants $c_k|_{k=1 ... 5}$ and $\hat{m}$ the 
un-renormalized heavy quark mass. The un-renormalized massive OME
is then obtained in Mellin space via the projection
  \begin{eqnarray}
    \label{eqc3}
     \Delta_T~\Ahathat_{qq,Q}^{\rm NS}\Bigl(\frac{\hat{m}^2}{\mu^2},\ep,N\Bigr)
     &=& 
         - i \frac{\delta^{ij}}{4N_c(\Delta.p)^{N+1} (D-2)}
           \Bigl\{
     {\sf Tr}[ \Delta\hspace*{-3mm}/~p\hspace*{-2mm}/~
               p^{\mu}\hat{G}^{ij, {\sf TR,NS}}_{\mu,q,Q}]
    -\Delta.p {\sf Tr}[p^{\mu}\hat{G}^{ij, {\sf TR,NS}}_{\mu,q,Q}] 
\N\\ && \hspace{40mm}
    +i\Delta.p {\sf Tr}[\sigma_{\mu \rho} p^\rho 
                        \hat{G}^{ij, {\rm TR,NS}}_{\mu,q,Q}]
          \Bigr\}~.
  \end{eqnarray}
Here $N_c$ denotes the number of colors. For the renormalization 
procedure and different steps to the final representation of the massive OME
in the $\MS$-scheme we refer to Ref.~\cite{BBK3}. Note that the 
renormalization  of the 
heavy quark mass is carried out in the on-mass-shell scheme.
\section{\boldmath The $O(a_s^2)$ Massive Operator Matrix Element}
\label{sec-3}

\vspace{1mm}\noindent
After mass renormalization the massive flavor non-singlet OME for transversity 
at $O(a_s^2)$ is given by \cite{BUZA1,BBK1} 
\begin{eqnarray}
\label{A2unren}
\Delta_T \hat{A}_{qq,Q}^{(2), \rm NS}(N) = S_\varepsilon^2 
\left(\frac{m^2}{\mu^2}\right)^\varepsilon \left\{\frac{1}{\varepsilon^2} 
\left[\beta_{0,Q}\gamma_{qq}^{(0),\rm TR}\right] + 
\frac{1}{\varepsilon} 
\left[
\frac{1}{2} \hat{\gamma}_{qq}^{(1),\rm TR} \right] 
+ a_{qq,Q}^{(2), \rm TR}
+ \varepsilon \overline{a}_{qq,Q}^{(2), \rm TR} \right\}~. 
\end{eqnarray}
Here we dropped all arguments on the r.h.s. $S_\varepsilon$ is the spherical
factor which occurs due to dimensional regularization and is set to one
in the $\MS$-scheme.
$\gamma_{qq}^{(k), \rm TR}(N)$ denote the $(k+1)$-loop anomalous 
dimensions for the non-singlet composite operator (\ref{op2}).
Note that as in Ref.~\cite{BBK3} we define the anomalous dimension 
corresponding to an operator $Z$-factor via
\begin{eqnarray}
\gamma = \mu \frac{\partial}{\partial \mu} \ln\left(Z(\mu)\right)~. 
\end{eqnarray}
$\beta_{0,Q}$ denotes the heavy flavor contribution to the $\beta$-function in 
lowest order,  
\begin{eqnarray}
\beta_{0,Q} &=&  - \frac{4}{3}~T_F~,  
\end{eqnarray}
with $T_F = 1/2$. Eq.~(\ref{A2unren}) has been expanded up to $O(\varepsilon)$ 
since 
the coefficient $\overline{a}_{qq,Q}^{(2),\rm TR}$ enters the 3-loop OME
via renormalization. The renormalized OME is given in Mellin space by 
  \begin{eqnarray}
     \Delta_T A_{qq,Q}^{(2),{\rm NS, \MS}}(N) &=&
                  \frac{\beta_{0,Q}\gamma_{qq}^{(0),\rm TR}}{4}
                    \ln^2 \Bigl(\frac{m^2}{\mu^2}\Bigr)
                 +\frac{\hat{\gamma}_{qq}^{(1), {\rm TR}}}{2}
                    \ln \Bigl(\frac{m^2}{\mu^2}\Bigr)
                 +a_{qq,Q}^{(2),{\rm TR}}
                 -\frac{\beta_{0,Q}\gamma_{qq}^{(0),\rm TR}}{4}\zeta_2 
                  \label{Aqq2NSTRQMSren} 
\end{eqnarray}
in the $\MS$-scheme and $\zeta_k,~k \geq 2, k \in {\bf N}$ denotes the 
Riemann 
$\zeta$-function at integer arguments. The calculation of the 2--loop OME in 
terms of Feynman-parameter integrals is straightforward, see \cite{BBK1}.
For the anomalous dimensions 
             $\gamma_{qq}^{(0), {\rm TR}}$ and 
$\hat{\gamma}_{qq}^{(1), {\rm TR}}$ we obtain
  \begin{eqnarray}
    \gamma_{qq}^{(0),{\rm TR}}(N) &=& 2 C_F \left[ -3 + 4 S_1\right],
   \label{gqqTR0}
\\
   \hat{\gamma}_{qq}^{(1), {\rm TR}}(N) &=& \frac{32}{9} C_F T_F \left[ 3 S_2
                                       - 5 S_1 + \frac{3}{8} \right]~,
   \label{gqqTR1}
  \end{eqnarray}
with $C_F = (N_c^2 - 1)/(2 N_c)$,  
confirming earlier results, 
  \cite{NLO1,NLO,NLO2}. Here, $S_k \equiv S_k(N)$ denote the single harmonic 
sums,  \cite{HSUM}. 
The finite and $O(\ep)$ contributions of the un-renormalized OME, 
Eq.~(\ref{A2unren}), read
  \begin{eqnarray}
   a_{qq,Q}^{(2), {\rm TR}}(N) &=& C_F T_F \Biggl\{ 
    -\frac{8}{3}    S_3
    +\frac{40}{9}   S_2
    -\left[ \frac{224}{27} 
           +\frac{8}{3}\zeta_2 \right] S_1
    +2 \zeta_2
    +{\frac{ \left( 24+73\,N+73\,{N}^{2} \right)}{18 N \left( N+1 \right) }}
    \Biggr\}~,   \N\\ \label{aqqTR2} 
\end{eqnarray}\begin{eqnarray}
   \overline{a}_{qq,Q}^{(2), {\rm TR}}(N) &=& C_F T_F \Biggl\{
      - \left[
               {\frac {656}{81}}\, 
              +{\frac {20}{9}}\, \zeta_2
              +{\frac {8}{9}}\, \zeta_3 \right] S_1
        +\left[{\frac {112}{27}}\, +\frac{4}{3}\, \zeta_2 \right] S_2
        -{\frac {20}{9}}\, S_3
         \N\\ &&
         +\frac{4}{3}\, S_4
         +\frac{1}{6}\, \zeta_2
         +\frac{2}{3}\, \zeta_3
      +\frac{ 
        \left( -144-48\,N+757\,{N}^{2}+1034\,{N}^{3}+517\,{N}^{4} \right) }
              {216 {N}^{2} \left( N+1 \right) ^{2}}
        \Biggr\}~. \N\\
            \label{aqqTR2bar}
  \end{eqnarray}
  The renormalized $2$--loop massive OME (\ref{Aqq2NSTRQMSren}) then becomes
  \begin{eqnarray}
\label{eq:A2TRN}
   \Delta_T A_{qq,Q}^{(2),{\rm NS, \MS}}(N) &=& C_F T_F \Biggl\{\left[
    -\frac{8}{3} S_1 + 2 \right] \ln^2\left(\frac{m^2}{\mu^2}\right)
    + \left[-{\frac {80}{9}} S_1 + \frac{2}{3} +\frac{16}{3} S_2
    \right] \ln\left(\frac{m^2}{\mu^2}\right)
    \N\\ && \hspace*{1cm}
    - \frac{8}{3}    S_3
    + \frac{40}{9}   S_2 
    - \frac{224}{27} S_1
    + \frac {24+73N+73{N}^{2}}{18 N \left( N+1\right) }    
     \Biggr\}~.  
  \end{eqnarray}
Corresponding quantities for vector currents were calculated in 
Refs.~\cite{BUZA1,BBK1}. In the limit $N \rightarrow \infty$
$\gamma_{qq}^{(0)}(N), 
 \hat{\gamma}_{qq}^{(1)}(N), 
a_{qq,Q}^{(2)}(N), \overline{a}_{qq,Q}^{(2)}(N)$
and $A_{qq,Q}^{(2)}(N)$
in the vector and transversity case approach each other.
This has also been observed for the 2-loop transversity anomalous dimension 
in Ref.~\cite{NLO}.
\section{\boldmath The $O(a_s^3)$ Massive Operator Matrix Element}
\label{sec-4}

\vspace{1mm}\noindent
The renormalized OME for transversity at $O(a_s^3)$ has the same structure
as the flavor non-singlet OME in the case of vector currents, \cite{BBK3}.
In Mellin space it is given by
  \begin{eqnarray}
    \Delta_T    A_{qq,Q}^{(3),{\rm NS}, \MS}(N)  &=&
     -\frac{\gamma_{qq}^{(0),{\rm TR}}\beta_{0,Q}}{6}
          \Bigl(
                 \beta_0
                +2\beta_{0,Q}
          \Bigr)
             \ln^3 \Bigl(\frac{m^2}{\mu^2}\Bigr)
         +\frac{1}{4}
          \Biggl\{
                   2\gamma_{qq}^{(1),{\rm TR}}\beta_{0,Q}\N
\N\\ 
&&
                  -2\hat{\gamma}_{qq}^{(1),{\rm TR}}
                             \Bigl(
                                    \beta_0
                                   +\beta_{0,Q}
                             \Bigr)
                  +\beta_{1,Q}\gamma_{qq}^{(0),{\rm TR}}
          \Biggr\}
             \ln^2 \Bigl(\frac{m^2}{\mu^2}\Bigr)
         +\frac{1}{2}
          \Biggl\{
                   \hat{\gamma}_{qq}^{(2),{\rm TR}}
\N\\ &&
                  -\Bigl(
                           4a_{qq,Q}^{(2),{\rm TR}}
                          -\zeta_2\beta_{0,Q}\gamma_{qq}^{(0),{\rm TR}}
                                    \Bigr)(\beta_0+\beta_{0,Q})
                  +\gamma_{qq}^{(0),{\rm TR}}\beta_{1,Q}^{(1)}
          \Biggr\}
             \ln \Bigl(\frac{m^2}{\mu^2}\Bigr)
\N\\&&
         +4\overline{a}_{qq,Q}^{(2),{\rm TR}}(\beta_0+\beta_{0,Q})
         -\gamma_{qq}^{(0)}\beta_{1,Q}^{(2)}
         -\frac{\gamma_{qq}^{(0),{\rm TR}}\beta_0\beta_{0,Q}\zeta_3}{6}
         -\frac{\gamma_{qq}^{(1),{\rm TR}}\beta_{0,Q}\zeta_2}{4}
\N\\ \N \\&&
         +2 \delta m_1^{(1)} \beta_{0,Q} \gamma_{qq}^{(0),{\rm TR}}
         +\delta m_1^{(0)} \hat{\gamma}_{qq}^{(1),{\rm TR}}
         +2 \delta m_1^{(-1)} a_{qq,Q}^{(2),{\rm TR}}
         +a_{qq,Q}^{(3),{\rm TR}} \label{Aqq3NSTRQMSren}
  \end{eqnarray}
in the $\MS$-scheme, performing  mass renormalization in the on-mass-shell 
scheme. Here the expansion coefficients of the $\beta$-function 
and the mass renormalization constants are, cf.~\cite{GWP,BBK3,TNW},
\begin{eqnarray}
\beta_0           &=& \frac{11}{3} C_A - \frac{4}{3} T_F N_f~,\\
\beta_{1,Q}       &=& -4 \left(\frac{5}{3} C_A + C_F\right) T_F~, \\
\beta_{1,Q}^{(1)} &=& -\frac{32}{9} T_F C_A + 15 T_F C_F~,   \\
\beta_{1,Q}^{(2)} &=& -\frac{86}{27} T_F C_A - \frac{31}{4} T_F C_f
- \zeta_2 T_F\left(\frac{5}{3} C_A + C_F\right)~, 
\\
\delta m_1^{(-1)}   &=& 6 C_F~,\\
\delta m_1^{(0)}    &=& -4 C_F~,\\
\delta m_1^{(1)}    &=& \left(4 + \frac{3}{4} \zeta_2 \right) C_F~,
\end{eqnarray}
with $C_A = N_c$, and the NLO anomalous dimension $\gamma_{qq}^{(1), \rm TR}$ 
reads, cf.~\cite{NLO1,NLO,NLO2},
\begin{eqnarray}
\gamma_{qq}^{(1), \rm TR}(N) &=&  
C_F^2 \left(4 S_2 -8 S_1 -1 \right)\N\\ &&
+ 8 C_F\left(C_F - \frac{C_A}{2}\right) \Biggl[-4 S_1 S_2
-8 S_1 S_{-2} + S_1 
- 4 S_3 - 4 S_{-3} + \frac{5}{2} S_2 \N\\ && \hspace*{4cm} + 8 S_{-2,1}
- \frac{1 -(-1)^{N}}{N(N+1)}  - \frac{1}{4} 
\Biggr] 
\N\\ &&
+ C_F C_A \left( -16 S_1 S_2 - \frac {58}{3} S_2 + \frac{572}{9} S_1
 -\frac {20}{3} \right) \N\\ &&
+ C_F T_F N_f \left( {\frac {32}{3}} S_2 -{\frac {160}{9}} S_1
+ \frac{4}{3}\right)~.
\end{eqnarray}

All contributions to Eq.~(\ref{Aqq3NSTRQMSren}) are known for general values 
of $N$, except of $\hat{\gamma}_{qq}^{(2),\rm TR}$ and 
$a_{qq,Q}^{(3),\rm TR}$, the constant contribution to the un-renormalized 
3-loop massive OME.
 Similarly to the vector case, Ref.~\cite{BBK3},
we calculate $\Delta_T    A_{qq,Q}^{(3),{\rm NS}, \MS}(N)$ for a fixed number 
of Mellin moments. The Feynman diagrams were generated by a code based on {\tt 
QGRAF} \cite{QGRAF} and the color algebra was performed using {\tt color.h} 
\cite{COLOR}. The computation is based on {\tt FORM} \cite{FORM} codes 
using ${\tt MATAD}$~\cite{MATAD}.
Since the projector in Eq.~(\ref{eqc3}) has to be applied we can calculate
the moments $N = 1$ to 13, i.e. one moment less than in the vector case in 
Ref.~\cite{BBK3}, given the complexity of the problem and the computer 
resources presently available. The computation time amounted to about 9 days.
The contributions to the 3-loop transversity  
anomalous dimension, $\hat{\gamma}_{qq}^{(2),\rm TR}(N)$, and the constant 
part 
of the un-renormalized massive transversity OME $a_{qq,Q}^{(3),\rm TR}(N)$ are 
given in the 
Appendix. In Figure~\ref{fig:1} the numerical values of $a_{qq,Q}^{(3),\rm 
TR}(N)$ are compared to those of $a_{qq,Q}^{(3),\rm V}(N)$, Ref.~\cite{BBK3}.
As has been observed at $O(a_s^2)$ already, for larger values of $N$ both
quantities approach each other at $O(a_s^3)$. This also applies to 
$\hat{\gamma}_{qq}^{(2),\rm (V,TR)}(N)$.

The present calculation confirms the $T_F$-parts of the transversity 3-loop 
anomalous dimension, which was calculated for $N = 1$ to 8 in 
Refs.~\cite{GRAC} for the first time. We also present the moments $N = 9 $ 
to 13. As a by-product of the present calculation the complete NLO anomalous 
dimension \cite{NLO,NLO1,NLO2} is confirmed for the moments $N = 1$ to 13.

  Finally, we show as examples the first moments of the
  $\overline{\sf MS}$--renormalized 
  $O(a_s^3)$ massive transversity OME. Unlike the case for
  the vector current, the first moment does not vanish, since there is no
  conservation law to enforce this. One obtains
    \begin{eqnarray}
     \Delta_T A_{qq,Q}^{(3),{\sf NS},\MS}(1)&=&C_FT_F\Biggl\{
         \Bigl(
              \frac{44}{27}C_A
             -\frac{16}{27}T_F(N_f+2)
         \Bigr)\ln^3\Bigl(\frac{m^2}{\mu^2}\Bigr)
        +\Bigl(
             \frac{32}{3}C_F
            -\frac{106}{9}C_A
\N\\&&\hspace{-25mm}
            -\frac{104}{27}T_F
         \Bigr)\ln^2\Bigl(\frac{m^2}{\mu^2}\Bigr)
        +\Biggl[
             \Bigl(
                    \frac{233}{9}
                   +16\zeta_3
             \Bigr)C_F
            +\Bigl(
                   -\frac{2233}{81}
                   -16\zeta_3
             \Bigr)C_A
            -\frac{604}{81}N_fT_F
\N\\&&\hspace{-25mm}
            -\frac{496}{81}T_F
         \Biggr]\ln\Bigl(\frac{m^2}{\mu^2}\Bigr)
         +\Bigl(
                -\frac{16}{3}B_4
                +24\zeta_4
                -\frac{278}{9}\zeta_3
                +\frac{7511}{81}
          \Bigr)C_F
         +\Bigl(
                 \frac{8}{3}B_4
                -24\zeta_4
\N\\&&\hspace{-25mm}
                +\frac{437}{27}\zeta_3
                -\frac{34135}{729}
          \Bigr)C_A
        +\Bigl(
               -\frac{6556}{729}
               +\frac{128}{27}\zeta_3
         \Bigr)T_FN_f
        +\Bigl(
                \frac{2746}{729}
               -\frac{224}{27}\zeta_3
         \Bigr)T_F
              \Biggr\}~,\\
     \Delta_T A_{qq,Q}^{(3),{\sf NS}, \MS}(2)&=&C_FT_F\Biggl\{
         \Bigl(
                \frac{44}{9}C_A
               -\frac{16}{9}T_F(N_f+2)
         \Bigr)\ln^3\Bigl(\frac{m^2}{\mu^2}\Bigr)
        +\Bigl(
               -\frac{34}{3}C_A
\N\\&&\hspace{-25mm}
               -8T_F
         \Bigr)\ln^2\Bigl(\frac{m^2}{\mu^2}\Bigr)
        +\Bigl[
              \Bigl(
                     15
                    +48\zeta_3
              \Bigr)C_F
             +\Bigl(
                    -\frac{73}{9}
                    -48\zeta_3
              \Bigr)C_A
               -\frac{196}{9}N_fT_F
\N\\&&\hspace{-25mm}
               -\frac{496}{27}T_F
         \Bigr]\ln\Bigl(\frac{m^2}{\mu^2}\Bigr)
       +\Bigl(
              -16B_4
              +72\zeta_4
              -\frac{310}{3}\zeta_3
              +\frac{4133}{27}
        \Bigr)C_F
       +\Bigl(
               8B_4
              -72\zeta_4
\N\\&&\hspace{-25mm}
              +\frac{533}{9}\zeta_3
              -56
        \Bigr)C_A
       +\Bigl(
              -\frac{1988}{81}
              +\frac{128}{9}\zeta_3
        \Bigr)T_FN_f
       +\Bigl(
               \frac{338}{27}
              -\frac{224}{9}\zeta_3
        \Bigr)T_F
                     \Biggr\}~,
\end{eqnarray}
with
   \begin{eqnarray}
\label{eqB4}
          {B_4}&=&-4\zeta_2\ln^2(2) +\frac{2}{3}\ln^4(2)
-\frac{13}{2}\zeta_4
                  +16 {\Li}_4\Bigl(\frac{1}{2}\Bigr) \label{B4} \\
                  &\approx&  -1.7628000871~. \nonumber
   \end{eqnarray}
  The structure of the massive OME is similar to the result for
  the unpolarized case, cf. Ref.~\cite{BBK3}, Eq.~(5.57). We checked the 
  moments 
  $N = 1$ to 4 keeping the complete dependence on the gauge--parameter $\xi$
  and find that it cancels in the final result. We observe that 
  the massive OME do not depend on $\zeta_2$~\footnote{The combination of 
multiple zeta values $B_4$ is characteristic for quantities depending on a 
single mass scale. In this specific combination $\zeta$-values at even 
integer argument contribute.}, as is also the case for 
  the various massive OMEs which were calculated for vector currents in 
  Ref.~\cite{BBK3}. The results for the massive OME for the moments $N = 1$ 
  to 13 and the quantities listed in the Appendix are attached to this paper 
  in {\tt FORM}-format.

  Since the light flavor Wilson coefficients for the processes from which the 
  transversity distribution can be extracted are not known 
  to 2-- and 3--loop order, phenomenological studies on  the effect of the 
  heavy flavor contributions cannot yet be performed. However, our
  results can be used in comparisons with 
  upcoming lattice simulations of operator matrix elements with 
  (2+1+1)-dynamical fermions including the charm quark. 
\section{\boldmath Remarks on the Soffer Bound}
\label{sec-5}

\vspace{1mm}\noindent
If the Soffer inequality \cite{SOFFER}
\begin{eqnarray}
\left|\Delta_T f(x,Q^2_0)\right|
\leq \frac{1}{2} \left[ f(x,Q^2_0) + \Delta f(x,Q^2_0)\right]
\label{SOF1}
\end{eqnarray}
holds for the non-perturbative parton distribution functions at a given scale
$Q_0^2$
one may check its generalization at the level of the corresponding
structure functions. In the light-flavor case this has been investigated
to $O(a_s)$ for the Drell-Yan process in Ref.~\cite{NLO1}. For the heavy
flavor corrections studied in the present paper one investigates
\begin{eqnarray}
\left|\Delta_T F(x,Q^2)\right|
\leq \frac{1}{2} \left[ F(x,Q^2) + \Delta F(x,Q^2)\right]~,
\label{SOF2}
\end{eqnarray}
where the structure functions are given in 
Eqs.~(\ref{eqstr1},\ref{eqstr2},\ref{eqWIL1})
and by corresponding relations in the longitudinally polarized case. One may
try to separate the evolution effects in the parton distribution functions
from those of the Wilson coefficients. 

\begin{figure}[htb]
\begin{center}
\includegraphics[angle=0, height=16.0cm]{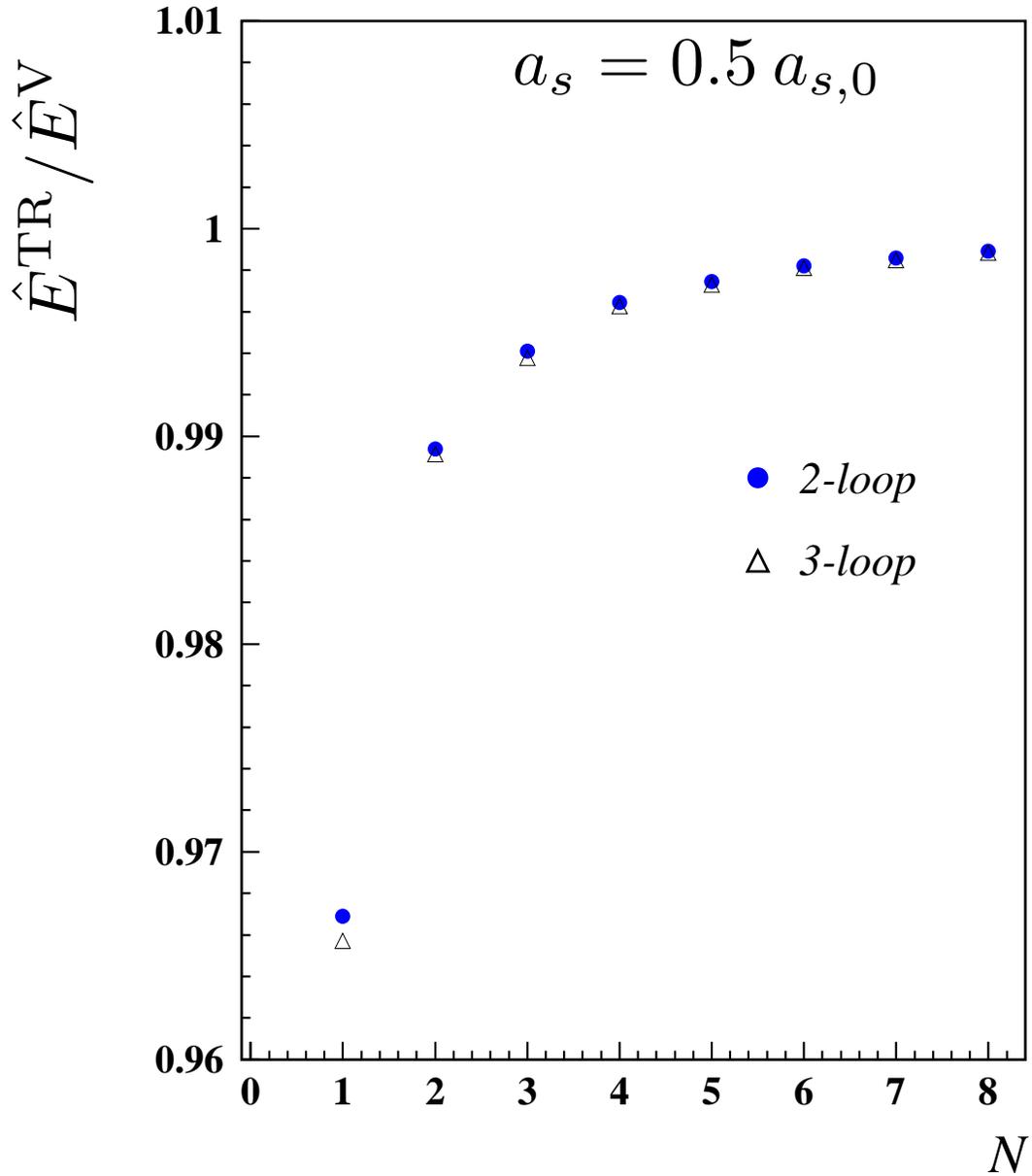}
\caption{\label{fig:2}
Ratio of the evolution operators $\hat{E}^{\rm TR,V}(N)$, 
Eq.~(\ref{eq:EVOL}), expanded up to the $O(a_s)$ terms (2 loops) and the 
$O(a_s^2)$
terms (3 loops), respectively, as a function of the Mellin variable $N$, with 
$\alpha_{s,0} = 0.3$.}
\end{center}
\end{figure}

The solution of the non-singlet evolution equation for the parton distribution 
$f^{\rm NS}(N,Q^2_0)$
in Mellin space for $N_f$ massless flavors reads to 3-loop order,
cf.~\cite{BBG}, 
\begin{eqnarray}
\label{eq:EVOL}
 f^{\rm NS}(N,Q^2) &=& E(N,Q^2,Q_0^2)~~f^{\rm NS}(N,Q^2_0) = 
\left(\frac{a}{a_0}\right)^{\gamma_{qq}^{(0),\rm NS}(N)/\beta_0} 
\hat{E}(N,Q^2,Q_0^2)~~f^{\rm NS}(N,Q^2_0) \N\\  
&=& 
\left(\frac{a}{a_0}\right)^{\gamma_{qq}^{(0),\rm NS}(N)/\beta_0}
\Biggl\{1 - \frac{1}{\beta_0} (a - a_0) \left[-\gamma_{qq}^{(1),\rm NS}(N)
+ \frac{\beta_1}{\beta_0} \gamma_{qq}^{(0),\rm NS}(N) \right]
\nonumber\\ &&
 - \frac{1}{2\beta_0}\left(a^2-a_0^2\right) \left[ - \gamma_{qq}^{(2),\rm 
NS}(N) + \frac{\beta_1}{\beta_0} 
\gamma_{qq}^{(1),\rm NS}(N)  - \left(
\frac{\beta_1^2}{\beta_0^2} - \frac{\beta_2}{\beta_0}\right) 
\gamma_{qq}^{(0),\rm NS}(N) \right]
\nonumber\\ && + \frac{1}{2\beta_0^2} (a-a_0)^2 \left(
\gamma_{qq}^{(1),\rm NS}(N) 
- \frac{\beta_1}{\beta_0} \gamma_{qq}^{(0),\rm NS}(N) \right)^2 \Biggr\} 
f^{\rm NS}(N,Q^2_0)~,
\end{eqnarray}
where $a_0 = a(Q_0^2)$ and
\begin{eqnarray}
\beta_1 &=& \frac{34}{3} C_A^2 - 4 C_F T_F N_f - \frac{20}{3} C_A T_F N_f \\
\beta_2 &=& \frac{2857}{54} C_A^3 + 2 C_F^2 T_F N_f - \frac{205}{9} C_F C_A
            T_F N_f 
            - \frac{1415}{27} C_A^2 T_F N_f + \frac{44}{9} C_F T_F^2 N_f^2
            + \frac{158}{27} C_A T_F^2 N_f^2~,
\nonumber\\
\end{eqnarray}
cf.~\cite{B12}. The moments of the anomalous dimensions for vector currents
are given in Refs.~\cite{GAMV}.
The evolution operator in the unpolarized and the longitudinally polarized 
case are the same due to a Ward identity. Therefore it is sufficient
to investigate the relation
\begin{eqnarray}
\left|\Delta_T E(N,Q^2)\right| \leq E^V(N,Q^2)~. 
\label{SOF3}
\end{eqnarray}
Up to the $O(a_s)$ corrections (NLO) the validity of this inequality was shown 
in \cite{NLO1}. Beyond this level only a finite number of Mellin moments can 
be compared for $\hat{E}^{\rm TR}(N,Q^2,Q_0^2)/\hat{E}^{\rm V}(N,Q^2,Q_0^2)$, 
for which the 3-loop transversity anomalous dimension is known \cite{GRAC}, 
expanding up to $O(a_s^2)$. This quantity is shown in Figure~\ref{fig:2} for 
the 2- and 3-loop case. The corresponding correction preserves the Soffer 
bound for characteristic values of $a_s$.

\begin{figure}[htb]
\begin{center}
\includegraphics[angle=0, height=16.0cm]{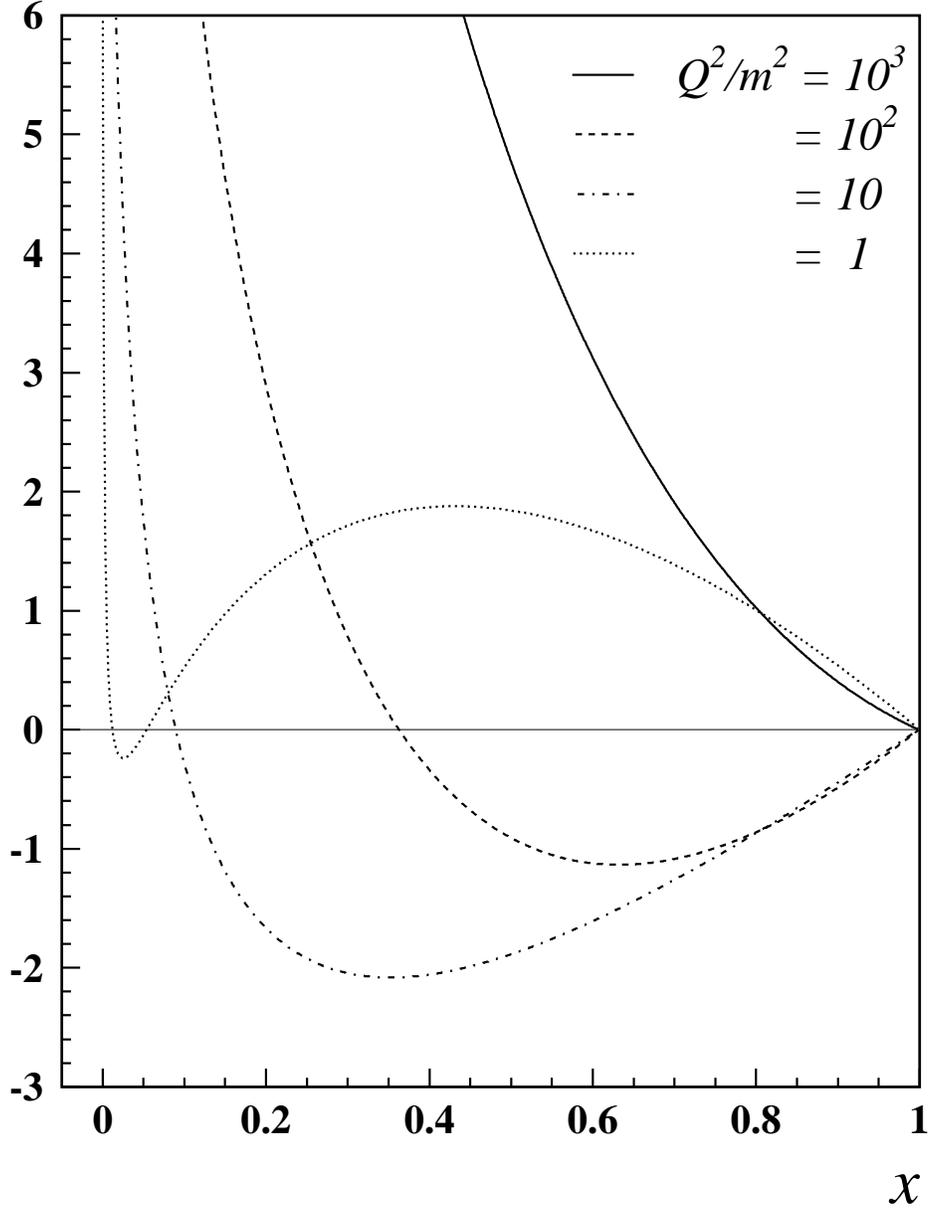}
\end{center}
\caption{\label{fig:3}
Difference of the massive OMEs $A_{qq,Q}^{(2),\rm NS, \MS}(x) - \Delta_T 
A_{qq,Q}^{(2),\rm  NS, \MS}(x)$, 
Eq.~(\ref{eq:A2dif}), for different ratios $Q^2/m^2$.}
\end{figure}

\begin{figure}[htb]
\begin{center}
\includegraphics[angle=0, height=16.0cm]{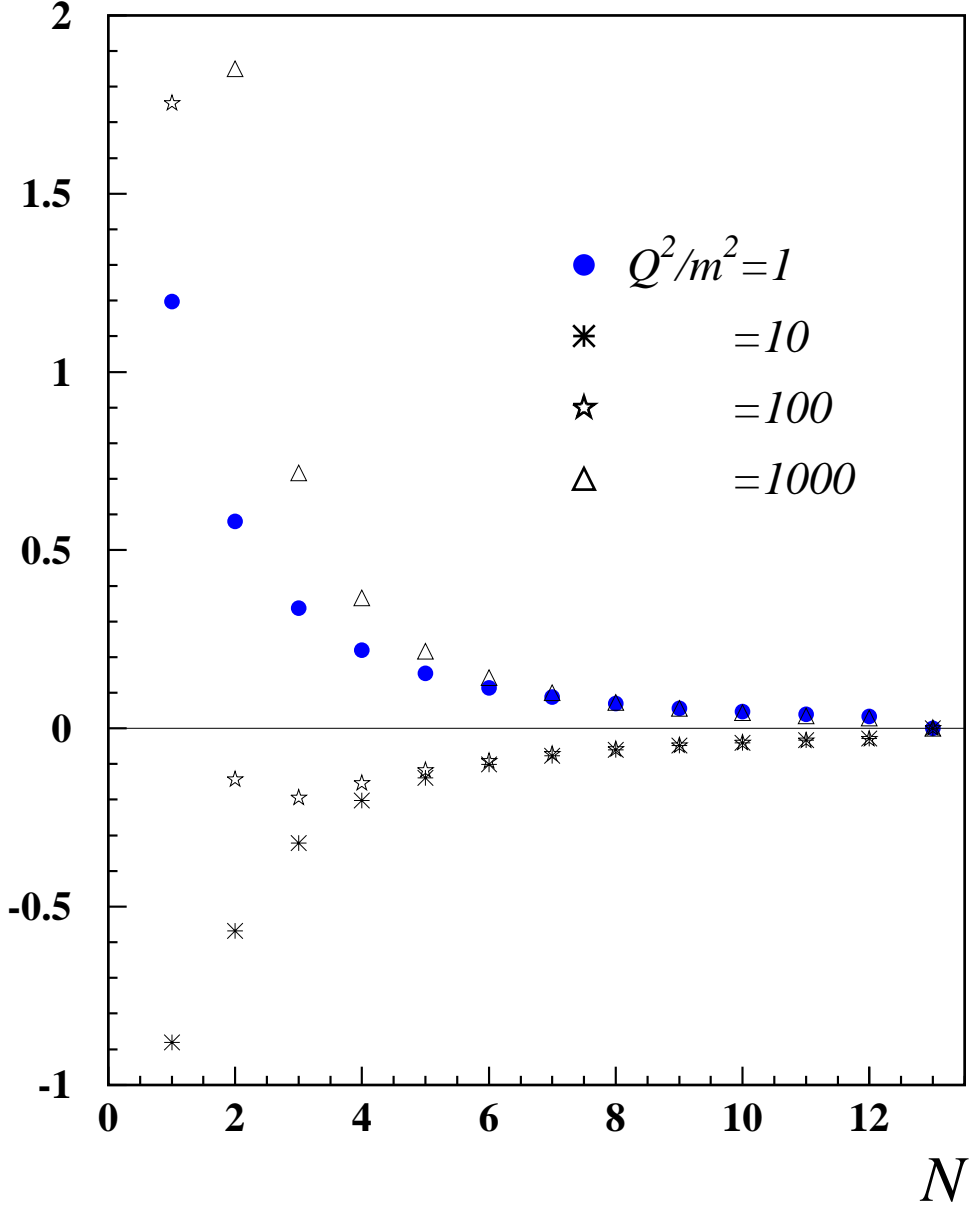}
\end{center}
\caption{\label{fig:4}
Difference of the massive OMEs $A_{qq,Q}^{(2),\rm NS, \MS}(N) - \Delta_T 
A_{qq,Q}^{(2),\rm NS, \MS}(N)$, 
Eq.~(\ref{eq:A2TRN}), and  Ref.~\cite{BUZA1},  Eq.~(3.35) for different ratios 
$Q^2/m^2$.}
\end{figure}

\begin{figure}[htb]
\begin{center}
\includegraphics[angle=0, height=16.0cm]{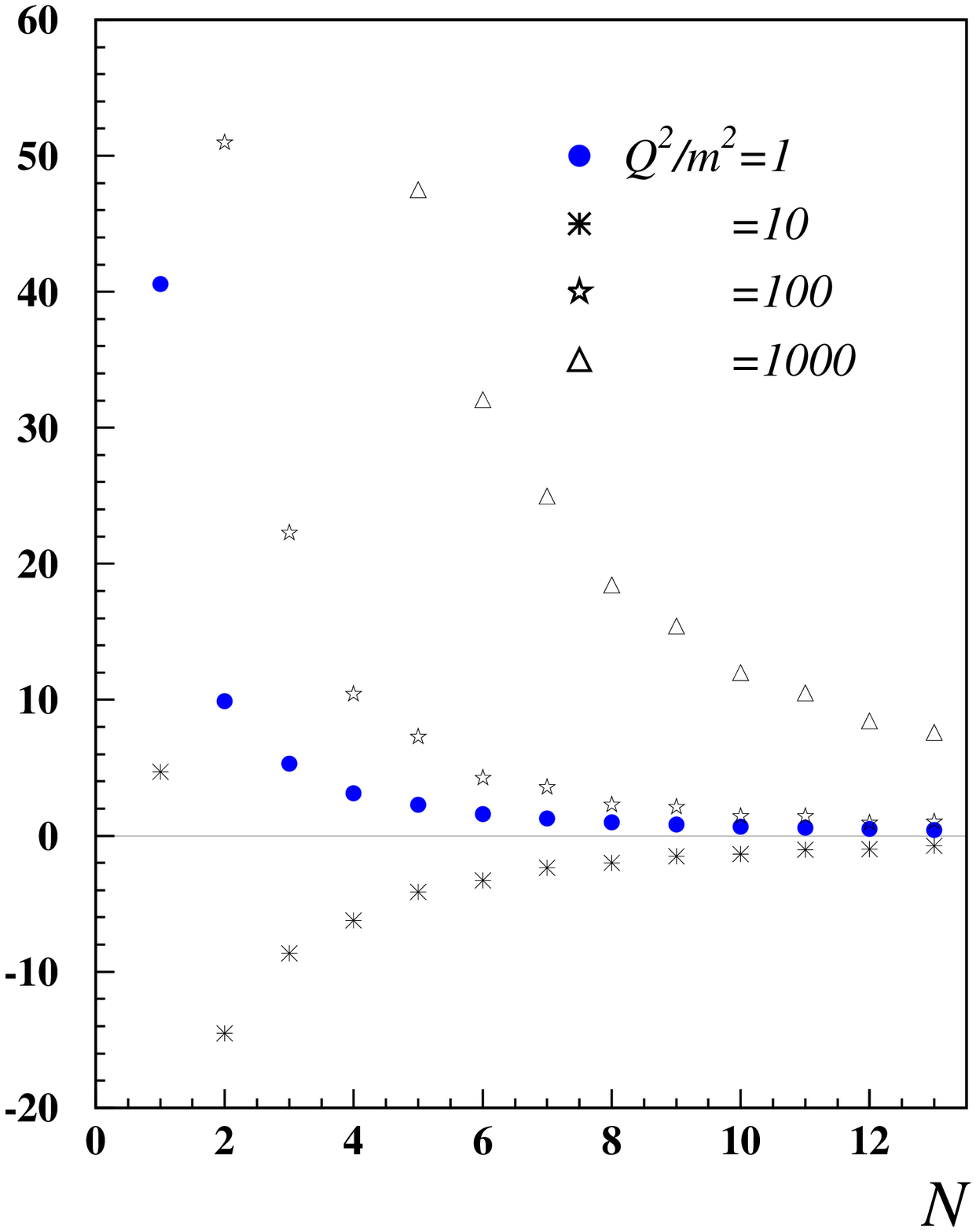}
\caption{Difference of the massive OMEs $A_{qq,Q}^{(3),\rm NS, \MS}(N) -
\Delta_T A_{qq,Q}^{(3),\rm NS, \MS}(N)$, 
Eq.~(\ref{Aqq3NSTRQMSren}), and Ref.~\cite{BBK3}, Eq.~(4.17) in the 
$\MS$-scheme for different ratios $Q^2/m^2$.}
\label{fig:5}
\end{center}
\end{figure}

Turning to the effect of the heavy flavor Wilson coefficient in the asymptotic 
region, Eq.~(\ref{HwcofTR}), we have to limit the investigation to the massive 
operator matrix elements since the corresponding light flavor Wilson 
coefficients were not yet calculated. In Figure~\ref{fig:3} we show the 
difference
\begin{eqnarray}
\label{eq:A2dif}
A_{qq,Q}^{(2),\rm NS, \MS}(x) - \Delta_T A_{qq,Q}^{(2), \rm NS, \MS}(x)
&=& C_F T_F (1-x) \Biggl \{\frac{4}{3}  
\ln^2\left(\frac{m^2}{\mu^2}\right)
                 + \frac{8}{3} \left(\ln(x) + \frac{11}{3}\right)
                \ln\left(\frac{m^2}{\mu^2}\right)
\nonumber \\ && \hspace*{2.5cm}
+ \frac{2}{3} \left[\ln^2(x) + \frac{22}{3} \ln(x) +
  \frac{116}{9}\right] \Biggr\}
\end{eqnarray}
for a series of values $Q^2 = \mu^2$. At large scales $A_{qq,Q}^{(2), \rm 
NS, \MS}(x) 
- \Delta_T A_{qq,Q}^{(2), \rm NS, \MS}(x)$ is positive and descending towards 
$x \simeq 1$, 
while 
at lower scales also negative values are reached in the intermediate region of 
$x$.
The difference is always positive in the small $x$ region. To maintain the 
Soffer bound the light flavor Wilson coefficients have to compensate the 
negative contributions. In Figure~\ref{fig:4} we show 
$A_{qq,Q}^{(2), \rm NS, \MS} 
- \Delta_T A_{qq,Q}^{(2), \rm NS, \MS}$ in Mellin space, where also a sign 
change is 
obtained. The behaviour of 
$A_{qq,Q}^{(3), \rm NS, \MS} - \Delta_T A_{qq,Q}^{(3), \rm NS, \MS}$, 
Figure~\ref{fig:5}, is quite 
similar to that shown in Figure~\ref{fig:4}
and a corresponding behaviour of $A_{qq,Q}^{(3), \rm NS, \MS}(x) - 
\Delta_T A_{qq,Q}^{(3), \rm NS, \MS}(x)$ to the one found at $O(a_s^2)$ is 
expected.
From the knowledge of the massive OMEs alone a conclusion on the 
validity of the Soffer bound at the level of the structure functions can not 
be drawn before the light flavor Wilson coefficients have been computed.
\section{\boldmath Conclusions}
\label{sec-6}

\vspace{1mm}\noindent
We calculated the flavor non-singlet massive OME for transversity at 2-loop 
order and for the Mellin moments $N = 1$ to 13 at 3-loop order. For large 
scales $Q^2 \gg m^2$ the heavy flavor Wilson coefficient can be determined 
from the light flavor Wilson coefficients and the respective 
process independent massive operator matrix element computed in the present 
paper. For flavor non-singlet quantities the heavy flavor corrections start 
at $O(a_s^2)$. The measurement of the corresponding scattering cross sections
requires high luminosity. In the present calculation we have verified the 
$T_F$-parts of the 3-loop transversity anomalous dimension for the moments 
$N=1$ to $8$ and extended this part up to $N=13$. As a general observation we
found that both the anomalous dimension and the expansion coefficients 
in $\varepsilon$ computed in the present calculation for transversity approach 
those in the vector case for large values of the Mellin parameter $N$.
We investigated the compatibility of the results of the present calculation
with the Soffer bound on the level of structure functions. While for the
evolution operator the Soffer bound is obeyed to 3-loop order, a final 
conclusion cannot be drawn for the massive operator matrix element at 
$O(a_s^2)$ and $O(a_s^3)$ alone concerning the massive Wilson 
coefficients
for the whole phase space, due to a sign change 
for $A_{qq,Q}^{\rm NS, \MS} - \Delta_T A_{qq,Q}^{\rm NS, \MS}$ at lower scales 
of $Q^2$ and 
medium values of $x$. A firm conclusion can only be drawn after the yet
unknown massless Wilson coefficients have been computed.

\vspace{5mm}\noindent
{\bf Acknowledgment.}~
We would like to thank I. Bierenbaum for discussions and her contributions to 
parts of the programs used in the present calculation and M. Steinhauser for 
providing the code {\tt MATAD 3.0}. This work was supported 
in part by DFG Sonderforschungsbereich Transregio 9, Computergest\"utzte 
Theoretische Teilchenphysik, Studienstiftung des 
Deutschen Volkes, and the European Commission MRTN HEPTOOLS under Contract No.
MRTN-CT-2006-035505.

\newpage
\section{Appendix}

\vspace{1mm}\noindent
The $T_F$--contributions to the 3-loop anomalous dimensions for $N=1$ to 13 
are given by~:
\begin{eqnarray}
\hat{\gamma}_{qq}^{(2),\rm TR}(1) &=& C_F T_F \Biggl[
-\frac{8}{3}\,T_F (2N_f+1)\,-{\frac {2008}{27}}\,{C_A}
+{\frac {196}{9}}\,{C_F}+32\,(C_F-C_A) {\zeta_3}  
\Biggr]\\
\hat{\gamma}_{qq}^{(2),\rm TR}(2) &=& C_F T_F \Biggl[
-{\frac {184}{27}}\, T_F ( 2 N_f+1)
-{\frac {2084}{27}}\,{C_A}-60\,
{C_F}+96 (C_F - C_A) \zeta_3
\Biggr]
\\
\hat{\gamma}_{qq}^{(2),\rm TR}(3) &=& C_F T_F \Biggl[
-{\frac {2408}{243}} T_F (2N_f+1)
-{\frac {19450}{243}}\,{C_A}
-{\frac {25276}{243}}\,{C_F}
+{\frac {416}{3}}\,(C_F-C_A) \zeta_3
\Biggr]
\\
\hat{\gamma}_{qq}^{(2),\rm TR}(4) &=& C_F T_F \Biggl[
-{\frac {14722}{1215}}\, T_F (2N_f+1)
-{\frac {199723}{2430}}\,{C_A}
-{\frac {66443}{486}}\, C_F
+{\frac {512}{3}} (C_F - C_A) \zeta_3
\Biggr]
\nonumber\\
\\
\hat{\gamma}_{qq}^{(2),\rm TR}(5) &=& C_F T_F \Biggl[
-{\frac {418594}{30375}}\,T_F (2N_f+1)
-{\frac {5113951}{60750}}\,{C_A}
-{\frac {49495163}{303750}}\,{C_F}
\nonumber\\ &&
+{\frac{2944}{15}}\, (C_F-C_A){\zeta_3}
\Biggr]
\\
\hat{\gamma}_{qq}^{(2),\rm TR}(6) &=& C_F T_F \Biggl[
-{\frac {3209758}{212625}}\,T_F(2N_f+1)
-{\frac {3682664}{42525}}\,{C_A}
-{\frac {18622301}{101250}}\,{C_F}
\nonumber\\ &&
+{\frac {1088}{5}}\,(C_F-C_A) \zeta_3
\Biggr]
\\
\hat{\gamma}_{qq}^{(2),\rm TR}(7) &=& C_F T_F \Biggl[
-{\frac {168501142}{10418625}}\,T_F (2N_f+1)
-{\frac {1844723441}{20837250}}\,{C_A}
-{\frac {49282560541}{243101250}}\,{C_F}
\nonumber\\ &&
+{\frac {8256}{35}}\,(C_F-C_A) \zeta_3
\Biggr]
\\
\hat{\gamma}_{qq}^{(2),\rm TR}(8) &=& C_F T_F \Biggl[
-{\frac {711801943}{41674500}}\, T_F(2N_f+1)
-{\frac {6056338297}{66679200}}\,C_A
-{\frac {849420853541}{3889620000}}\,C_F
\nonumber\\ &&
+{\frac {8816}{35}}\,(C_F-C_A) \zeta_3
\Biggr]
\\
\hat{\gamma}_{qq}^{(2),\rm TR}(9) &=& C_F T_F \Biggl[
-{\frac {20096458061}{1125211500}}\, T_F (2N_f+1)
-{\frac {119131812533}{1285956000}}\,C_A
-{\frac {24479706761047}{105019740000}}\,C_F
\nonumber\\ &&
+{\frac {83824}{315}}\,(C_F-C_A)\zeta_3
\Biggr]
\\
\hat{\gamma}_{qq}^{(2),\rm TR}(10) &=& C_F T_F \Biggl[
-{\frac {229508848783}{12377326500}}\, T_F (2N_f+1)
-{\frac {4264058299021}{45008460000}}\, C_A
-{\frac {25800817445759}{105019740000}}\, C_F
\nonumber\\ &&
+{\frac {87856}{315}}\,(C_F-C_A) \zeta_3
\Biggr]
\end{eqnarray} \begin{eqnarray}
\hat{\gamma}_{qq}^{(2),\rm TR}(11) &=& C_F T_F \Biggl[
-{\frac {28677274464343}{1497656506500}}\, T_F (2N_f+1)
-{\frac {75010870835743}{778003380000}}\, C_A
\nonumber
\end{eqnarray}\begin{eqnarray}
&&
-{\frac {396383896707569599}{1537594013340000}}\, C_F
+{\frac {1006736}{3465}}\, (C_F-C_A) \zeta_3
\Biggr]
\\
\hat{\gamma}_{qq}^{(2),\rm TR}(12) &=& C_F T_F \Biggl[
-{\frac {383379490933459}{19469534584500}}\,{T_F} (2 N_f+1)
-{\frac {38283693844132279}{389390691690000}}\,{C_A}
\nonumber\\ &&
-{\frac {1237841854306528417}{4612782040020000}}\,{C_F}
+{\frac {1043696}{3465}}\, (C_F - C_A) \zeta_3
\Biggr]\\
\hat{\gamma}_{qq}^{(2),\rm TR}(13) &=& C_F T_F \Biggl[
-{\frac {66409807459266571}{3290351344780500}}\, T_F (2N_f+1)
-{\frac {6571493644375020121}{65807026895610000}}\, C_A
\nonumber\\ &&
-{\frac {36713319015407141570017}{131745667845011220000}}\, C_F
+{\frac {14011568}{45045}}\,(C_F-C_A) \zeta_3
\Biggr]~.
\end{eqnarray}
\newpage

\vspace{1mm}\noindent
The constant parts $a_{qq,Q}^{(3), \rm TR}(N)$ of the massive 3-loop 
OME for $N=1$ to 13 are given by~:
\begin{eqnarray}
a_{qq,Q}^{(3), \rm TR}(1) &=& C_F T_F \Biggl[
 \left( {\frac {481}{27}}\,{\zeta_3}+\frac{8}{3}\,{B_4}-24\,{\zeta_4}-{\frac 
{61}
{27}}\,{\zeta_2}-{\frac {26441}{1458}} \right) {C_A}
\nonumber\\ &&
+ \left( -{\frac {52}{27}}\,{\zeta_2}+{\frac {112}{27}}\,{\zeta_3}-{\frac {
15850}{729}} \right) {N_f}\,{T_F}
\nonumber\\ &&
+ \left( -{\frac {104}{27}}\,{\zeta_2}-{\frac {6548}{729}}
-{\frac {256}{27}}\,{\zeta_3} \right) {T_F}
\nonumber\\ &&
+ \left( -{\frac {278}{9}}\,{\zeta_3}+{\frac {49}{3}}
\,{\zeta_2}+{\frac {15715}{162}}-\frac{16}{3}\,{B_4}+24\,{\zeta_4} \right) 
{C_F}
\Biggr]
\\
a_{qq,Q}^{(3), \rm TR}(2) &=& C_F T_F \Biggl[
\left( {\frac 
{577}{9}}\,{\zeta_3}+8\,{B_4}-72\,{\zeta_4}+\frac{1}{3}\,{\zeta_2}+
{\frac {1043}{162}} \right){C_A} 
\nonumber\\ &&
+ \left( -4\,{\zeta_2}+{\frac {112}{9}}\,{\zeta_3}-{\frac {4390}{81}} \right) 
{N_f}\,{T_F}
\nonumber\\ &&
+ \left( -8\,{\zeta_2}-{\frac {1388}{81}}-{\frac {256}{9}}
\,{\zeta_3} \right) {T_F}
\nonumber\\ &&
+ \left( -{\frac {310}{3}}\,{\zeta_3}+33\,{\zeta_2}+{\frac {10255}{54}}-16\,{B_4}
+72\,{\zeta_4} \right) {C_F}
\Biggr]
\\
a_{qq,Q}^{(3), \rm TR}(3) &=& C_F T_F \Biggl[
\left( {\frac {40001}{405}}\,{\zeta_3}+{\frac {104}{9}}\,{B_4}-104\,{
\zeta_4}+{\frac {121}{81}}\,{\zeta_2}+{\frac {327967}{21870}} \right) 
{C_A}
\nonumber\\ &&
+ \left( -{\frac {452}{81}}\,{\zeta_2}+{\frac {1456}
{81}}\,{\zeta_3}-{\frac {168704}{2187}} \right) {N_f}\,{T_F}
\nonumber\\ &&
+ \left( -{\frac {904}{81}}\,{\zeta_2}-{\frac {52096}{2187}}-{\frac
{3328}{81}}\,{\zeta_3} \right) {T_F}
\nonumber\\ &&
+ \left( -{\frac {1354
}{9}}\,{\zeta_3}+{\frac {3821}{81}}\,{\zeta_2}+{\frac {1170943}{4374}}-{
\frac {208}{9}}\,{B_4}+104\,{\zeta_4} \right) C_F
\Biggr]
\\
a_{qq,Q}^{(3), \rm TR}(4) &=& C_F T_F \Biggl[
\left( {\frac {52112}{405}}\,{\zeta_3}+{\frac {128}{9}}\,{B_4}-128\,{
\zeta_4}+{\frac {250}{81}}\,{\zeta_2}+{\frac {4400353}{218700}} \right) 
{C_A}
\nonumber\\ &&
+ \left( -{\frac {554}{81}}\,{\zeta_2}+{\frac {
1792}{81}}\,{\zeta_3}-{\frac {20731907}{218700}} \right) 
{N_f}\,{T_F}
\nonumber\\ &&
+ \left( -{\frac {1108}{81}}\,{\zeta_2}-{\frac {3195707}{
109350}}-{\frac {4096}{81}}\,{\zeta_3} \right) 
{T_F}
\nonumber\\ &&
+ \left( -{\frac {556}{3}}\,{\zeta_3}+{\frac {4616}{81}}\,{\zeta_2}+{\frac {
56375659}{174960}}-{\frac {256}{9}}\,{B_4}+128\,{\zeta_4} \right) C_F
\Biggr]
\\
a_{qq,Q}^{(3), \rm TR}(5) &=& C_F T_F \Biggl[
\left( {\frac {442628}{2835}}\,{\zeta_3}+{\frac {736}{45}}\,{B_4}-{
\frac {736}{5}}\,{\zeta_4}+{\frac {8488}{2025}}\,{\zeta_2}+{\frac {
1436867309}{76545000}} \right) 
C_A
\nonumber\\ &&
+ \left( -{
\frac {15962}{2025}}\,{\zeta_2}+{\frac {10304}{405}}\,{\zeta_3}-{\frac {
596707139}{5467500}} \right) 
N_f T_F
\nonumber
\end{eqnarray}\begin{eqnarray}
&&
+ \left( -{
\frac {31924}{2025}}\,{\zeta_2}-{\frac {92220539}{2733750}}-{\frac {23552}{
405}}\,{\zeta_3} \right) 
T_F\nonumber\\
&&
+ \left( -{\frac {47932}{225
}}\,{\zeta_3}+{\frac {662674}{10125}}\,{\zeta_2}+{\frac {40410914719}{
109350000}}-{\frac {1472}{45}}\,{B_4}+{\frac {736}{5}}\,{\zeta_4}
 \right) 
C_F
\Biggr]
\\
a_{qq,Q}^{(3), \rm TR}(6) &=& C_F T_F \Biggl[
\left( {\frac {172138}{945}}\,{\zeta_3}+{\frac {272}{15}}\,{B_4}-{
\frac {816}{5}}\,{\zeta_4}+{\frac {10837}{2025}}\,{\zeta_2}+{\frac {
807041747}{53581500}} \right) 
C_A
\nonumber\\ &&
+ \left( -{
\frac {17762}{2025}}\,{\zeta_2}+{\frac {3808}{135}}\,{\zeta_3}-{\frac {
32472719011}{267907500}} \right) 
N_f T_F
\nonumber\\ &&
+ \left(
-{\frac {35524}{2025}}\,{\zeta_2}-{\frac {5036315611}{133953750}}-{\frac {
8704}{135}}\,{\zeta_3} \right) 
T_F
\nonumber\\ &&
+ \left( -{\frac {
159296}{675}}\,{\zeta_3}+{\frac {81181}{1125}}\,{\zeta_2}+{\frac {
14845987993}{36450000}}-{\frac {544}{15}}\,{B_4}+{\frac {816}{5}}\,{
\zeta_4} \right) 
C_F
\Biggr]
\\
a_{qq,Q}^{(3), \rm TR}(7) &=& C_F T_F \Biggl[
\left( {\frac {27982}{135}}\,{\zeta_3}+{\frac {688}{35}}\,{B_4}-{\frac
{6192}{35}}\,{\zeta_4}+{\frac {620686}{99225}}\,{\zeta_2}+{\frac {
413587780793}{52509870000}} \right) 
C_A
\nonumber\\ &&
+ \left( -
{\frac {947138}{99225}}\,{\zeta_2}+{\frac {1376}{45}}\,{\zeta_3}-{\frac {
1727972700289}{13127467500}} \right) 
N_f T_F
\nonumber\\ &&
+ \left( -{\frac {1894276}{99225}}\,{\zeta_2}-{\frac {268946573689}{
6563733750}}-{\frac {22016}{315}}\,{\zeta_3} \right) 
T_F
\nonumber\\ &&
+ \left( -{\frac {8454104}{33075}}\,{\zeta_3}+{\frac {90495089}{1157625}}\,
{\zeta_2}+{\frac {12873570421651}{29172150000}}-{\frac {1376}{35}}\,{B_4
}+{\frac {6192}{35}}\,{\zeta_4} \right) 
C_F
\Biggr]
\nonumber\\
\\
a_{qq,Q}^{(3), \rm TR}(8) &=& C_F T_F \Biggl[
\left( {\frac {87613}{378}}\,{\zeta_3}+{\frac {2204}{105}}\,{B_4}-{
\frac {6612}{35}}\,{\zeta_4}+{\frac {11372923}{1587600}}\,{\zeta_2}-{\frac {
91321974347}{112021056000}} \right) 
C_A
\nonumber\\ &&
+ \left( -
{\frac {2030251}{198450}}\,{\zeta_2}+{\frac {4408}{135}}\,{\zeta_3}-{\frac {
29573247248999}{210039480000}} \right) 
N_f T_F
\nonumber\\ &&
+ \left( -{\frac {2030251}{99225}}\,{\zeta_2}-{\frac {4618094363399}{
105019740000}}-{\frac {70528}{945}}\,{\zeta_3} \right) 
T_F
\nonumber\\ &&
+ \Biggl( -{\frac {9020054}{33075}}\,{\zeta_3}+{\frac 
{171321401}{2058000
}}\,{\zeta_2}+{\frac {1316283829306051}{2800526400000}}
\nonumber\\ &&
-{\frac {4408}{105}}
\,{B_4}+{\frac {6612}{35}}\,{\zeta_4} \Biggr) 
C_F
\Biggr]
\\
a_{qq,Q}^{(3), \rm TR}(9) &=& C_F T_F \Biggl[
\Biggl( {\frac {9574759}{37422}}\,{\zeta_3}+{\frac {20956}{945}}\,{B_4}-
{\frac {20956}{105}}\,{\zeta_4}+{\frac {16154189}{2041200}}\,{\zeta_2}
\nonumber\\ &&
-{
\frac {17524721583739067}{1497161413440000}} \Biggr) 
C_A
\nonumber\\ &&
+ \left( -{\frac {19369859}{1786050}}\,{\zeta_2}+{\frac {41912}{1215
}}\,{\zeta_3}-{\frac {2534665670688119}{17013197880000}} \right) 
N_f T_F
\N
\end{eqnarray}\begin{eqnarray}
&&
+ \left( -{\frac {19369859}{893025}}\,{\zeta_2}-{
\frac {397003835114519}{8506598940000}}-{\frac {670592}{8505}}\,{\zeta_3}
 \right) 
T_F
\nonumber\\ &&
+ \Biggl( -{\frac {85698286}{297675}}\,{\zeta_3}+{\frac 
{131876277049}{1500282000}}\,{\zeta_2}+{\frac {
1013649109952401819}{2041583745600000}}
\nonumber
\\
&& 
-{\frac {41912}{945}}\,{B_4}+{
\frac {20956}{105}}\,{\zeta_4} \Biggr) 
C_F
\Biggr]
\\
a_{qq,Q}^{(3), \rm TR}(10) &=& C_F T_F \Biggl[
\Biggl( {\frac {261607183}{935550}}\,{\zeta_3}+{\frac {21964}{945}}\,{B_4}
-{\frac {21964}{105}}\,{\zeta_4}+{\frac {618627019}{71442000}}\,{\zeta_2}
\nonumber\\ &&
-{\frac {176834434840947469}{7485807067200000}} \Biggr) 
C_A
\nonumber\\ &&
+ \left( -{\frac {4072951}{357210}}\,{\zeta_2}+{\frac {43928}{
1215}}\,{\zeta_3}-{\frac {321908083399769663}{2058596943480000}} \right) 
N_f T_F
\nonumber\\ &&
+ \left( -{\frac {4072951}{178605}}\,{\zeta_2}-{\frac 
{50558522757917663}{1029298471740000}}-{\frac {702848}{8505}}\,
{\zeta_3} \right) 
T_F
\nonumber\\ &&
+ \Biggl( -{\frac {3590290}{11907}}
\,{\zeta_3}+{\frac {137983320397}{1500282000}}\,{\zeta_2}+{\frac {
11669499797141374121}{22457421201600000}}
\nonumber\\ &&
-{\frac {43928}{945}}\,{B_4}+{
\frac {21964}{105}}\,{\zeta_4} \Biggr) 
C_F
\Biggr]
\\
a_{qq,Q}^{(3), \rm TR}(11) &=& C_F T_F \Biggl[
\Biggl( {\frac {3687221539}{12162150}}\,{\zeta_3}+{\frac {251684}{10395}}\,
{B_4}-{\frac {251684}{1155}}\,{\zeta_4}+{\frac {149112401}{16038000}}\,{
\zeta_2}
\nonumber\\ &&
-{\frac {436508000489627050837}{11775174516705600000}} \Biggr) 
C_A
\nonumber\\ &&
+ \left( -{\frac {514841791}{43222410}}\,{\zeta_2}+{\frac 
{503368}{13365}}\,{\zeta_3}-{\frac {40628987857774916423}{
249090230161080000}} \right) 
N_f T_F
\nonumber\\ &&
+ \left( -{
\frac {514841791}{21611205}}\,{\zeta_2}-{\frac {6396997235105384423}{
124545115080540000}}-{\frac {8053888}{93555}}\,{\zeta_3} \right) 
T_F
\nonumber\\ &&
+ \Biggl( -{\frac {452259130}{1440747}}\,{\zeta_3}+{\frac {
191230589104127}{1996875342000}}\,{\zeta_2}+{\frac {
177979311179110818909401}{328799103812625600000}}
\nonumber\\ &&
-{\frac {503368}{10395}}
\,{B_4}+{\frac {251684}{1155}}\,{\zeta_4} \Biggr) 
C_F
\Biggr]
\\
a_{qq,Q}^{(3), \rm TR}(12) &=& C_F T_F \Biggl[
 \Biggl( {\frac {85827712409}{8644482000}}\,{\zeta_2}
-{\frac {245210883820358086333}{4783664647411650000}}
+{\frac {260924}{10395}}\,{B_4}-{\frac{260924}{1155}}\,{\zeta_4}
\nonumber\\ &&
+{\frac {3971470819}{12162150}}\,{\zeta_3} \Biggr){C_A}
\nonumber\\ &&
+ \left( -{\frac {7126865031281296825487}{42096248897222520000}}
+{\frac {521848}{13365}}\,{\zeta_3}
-{\frac {535118971}{43222410}}\,{\zeta_2} \right) N_f\,{T_F}
\nonumber
\end{eqnarray}\begin{eqnarray}
&&
+ \left( -{\frac {8349568}{93555}}\,{\zeta_3}
-{\frac {535118971}{21611205}}\,{\zeta_2}
-{\frac {1124652164258976877487}{21048124448611260000}} \right) {T_F}
\nonumber\\ &&
+ \Biggl( {\frac {260924}{1155}}\,{\zeta_4}
+{\frac {2396383721714622551610173}{4274388349564132800000}}
-{\frac {468587596}{1440747}}\,{\zeta_3}
\nonumber\\ &&
-{\frac {521848}{10395}}\,{B_4}
+{\frac {198011292882437}{1996875342000}}\,{\zeta_2} \Biggr) {C_F}
\Biggr\}
\\
a_{qq,Q}^{(3), \rm TR}(13) &=& C_F T_F \Biggl[
 \Biggl( {\frac {15314434459241}{1460917458000}}\,{\zeta_2}
-{\frac {430633219615523278883051}{6467514603300550800000}}
+{\frac {3502892}{135135}}\,{B_4}
\nonumber\\ &&
-{\frac{3502892}{15015}}\,{\zeta_4}
+{\frac {327241423}{935550}}\,{\zeta_3} \Biggr){C_A}
\nonumber\\ &&
+ \left( -{\frac {1245167831299024242467303}{7114266063630605880000}}
+{\frac {7005784}{173745}}\,{\zeta_3}
-{\frac {93611152819}{7304587290}}\,{\zeta_2} \right) N_f\,{T_F}
\nonumber\\ &&
+ \left( -{\frac {112092544}{1216215}}\,{\zeta_3}
-{\frac {93611152819}{3652293645}}\,{\zeta_2}
-{\frac {196897887865971730295303}{3557133031815302940000}} \right) {T_F}
\nonumber\\ &&
+ \Biggl( {\frac {3502892}{15015}}\,{\zeta_4}
+{\frac {70680445585608577308861582893}{122080805651901196900800000}}
-{\frac {81735983092}{243486243}}\,{\zeta_3}
\nonumber\\ &&
-{\frac {7005784}{135135}}\,{B_4}
+{\frac {449066258795623169}{4387135126374000}}\,{\zeta_2} \Biggr) {C_F}
\Biggr\}~.
\end{eqnarray}

\newpage


\newpage
\begin{figure}[htb]
\begin{center}
\includegraphics[angle=0, height=16.0cm]{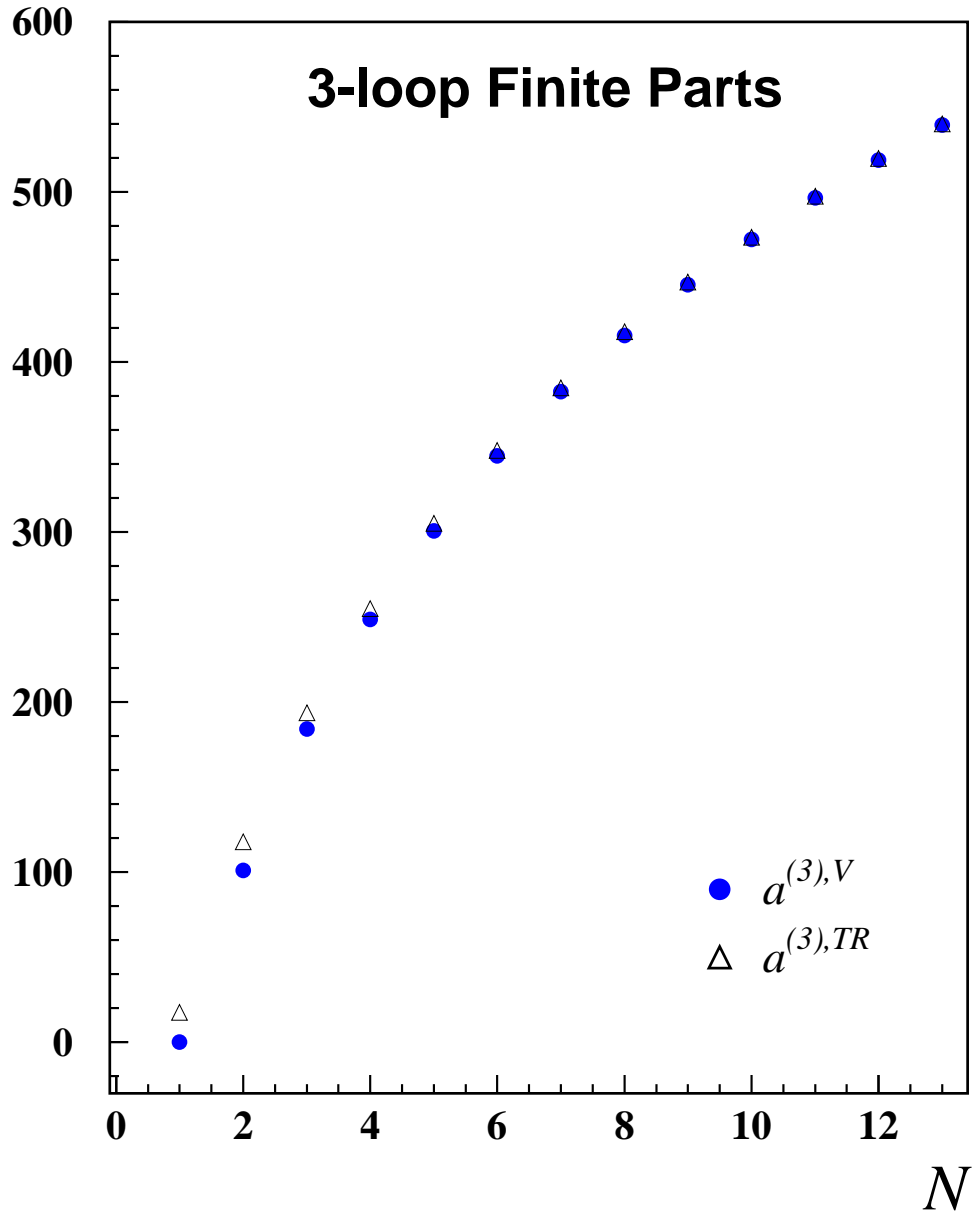}
\caption{\label{fig:1}
The constant part $a_{qq,Q}^{(3)}$ of the un-renormalized flavor non-singlet 
massive 3-loop OME 
in the vector case  \cite{BBK3} and for transversity for $N_f = 3$.} 
\end{center}
\end{figure}

\newpage

\newpage

\newpage

\newpage

\begin{thebibliography}{99}
%
\bibitem{FEYN}
R.P. Feynman, {\sf Photon hadron interactions}, (Benjamin, New York, 1972).
%
\bibitem{PROC1}
  J.~P.~Ralston and D.~E.~Soper,
  Nucl.\ Phys.\  B {\bf 152} (1979) 109;\\
  R.~L.~Jaffe and X.~D.~Ji,
  Phys.\ Rev.\ Lett.\  {\bf 67} (1991) 552;
  Nucl.\ Phys.\  B {\bf 375} (1992) 527.
%
\bibitem{PROC2}
  J.~L.~Cortes, B.~Pire and J.~P.~Ralston,
  Z.\ Phys.\  C {\bf 55} (1992) 409.
%
\bibitem{PROC3}
  J.~C.~Collins,
  Nucl.\ Phys.\  B {\bf 396} (1993) 161
  [arXiv:hep-ph/9208213];\\
  R.~L.~Jaffe and X.~D.~Ji,
  Phys.\ Rev.\ Lett.\  {\bf 71} (1993) 2547
  [arXiv:hep-ph/9307329];\\
  R.~D.~Tangerman and P.~J.~Mulders,
  arXiv:hep-ph/9408305;\\
  D.~Boer and P.~J.~Mulders,
  Phys.\ Rev.\  D {\bf 57} (1998) 5780
  [arXiv:hep-ph/9711485].
%
\bibitem{ARTR}
  X.~Artru and M.~Mekhfi,
  Z.\ Phys.\  C {\bf 45} (1990) 669.
%
\bibitem{RATCL}
  V.~Barone, A.~Drago and P.~G.~Ratcliffe,
  Phys.\ Rept.\  {\bf 359} (2002) 1
  [arXiv:hep-ph/0104283].
%
\bibitem{EXP}
  A.~Airapetian {\it et al.}  [HERMES Collaboration],
  Phys.\ Rev.\ Lett.\  {\bf 94} (2005) 012002
  [arXiv:hep-ex/0408013];\\
  A.~Airapetian {\it et al.}  [HERMES Collaboration],
  JHEP {\bf 0806} (2008) 017
  [arXiv:0803.2367 [hep-ex]];\\
  M.~Alekseev {\it et al.}  [COMPASS Collaboration],
  Phys.\ Lett.\  B {\bf 673} (2009) 127
  [arXiv:0802.2160 [hep-ex]];
  V.~Y.~Alexakhin {\it et al.}  [COMPASS Collaboration],
  Phys.\ Rev.\ Lett.\  {\bf 94} (2005) 202002
  [arXiv:hep-ex/0503002];\\
COMPASS collaboration, private communication.\\
  M.~F.~Lutz, B.~Pire, O.~Scholten and R.~Timmermans et al., [The PANDA
                  Collaboration],
  {\sf Physics Performance Report for PANDA: Strong Interaction Studies with
       Antiprotons},
  arXiv:0903.3905 [hep-ex];\\
  A.~Afanasev {\it et al.},
  arXiv:hep-ph/0703288.
%
\bibitem{PHEN}
cf. Section~8, Ref.~\cite{RATCL}.
%
\bibitem{ANSELM}
  M.~Anselmino, M.~Boglione, U.~D'Alesio, A.~Kotzinian, F.~Murgia, A.~Prokudin and C.~Turk,
  Phys.\ Rev.\  D {\bf 75} (2007) 054032
  [arXiv:hep-ph/0701006];\\
  M.~Anselmino, M.~Boglione, U.~D'Alesio, A.~Kotzinian, F.~Murgia, A.~Prokudin and S.~Melis,
  arXiv:0812.4366 [hep-ph].
%
\bibitem{LATT}
  S.~Aoki, M.~Doui, T.~Hatsuda and Y.~Kuramashi,
  Phys.\ Rev.\  D {\bf 56} (1997) 433
  [arXiv:hep-lat/9608115];\\
  M.~G\"ockeler {\it et al.},
  Nucl.\ Phys.\ Proc.\ Suppl.\  {\bf 53} (1997) 315
  [arXiv:hep-lat/9609039];\\
  A.~Ali~Khan {\it et al.},
  Nucl.\ Phys.\ Proc.\ Suppl.\  {\bf 140} (2005) 408
  [arXiv:hep-lat/0409161];\\
  D.~Dolgov {\it et al.}  [LHPC collaboration and TXL Collaboration],
  Phys.\ Rev.\  D {\bf 66} (2002) 034506
  [arXiv:hep-lat/0201021];\\
  M.~Diehl {\it et al.}  [QCDSF Collaboration and UKQCD Collaboration],
  arXiv:hep-ph/0511032;\\
  M.~G\"ockeler {\it et al.}  [QCDSF Collaboration and UKQCD Collaboration],
  Phys.\ Rev.\ Lett.\  {\bf 98} (2007) 222001
  [arXiv:hep-lat/0612032];\\
D. Renner, private communication.
%
\bibitem{LO}
  F.~Baldracchini, N.~S.~Craigie, V.~Roberto and M.~Socolovsky,
  Fortsch.\ Phys.\  {\bf 30} (1981) 505
  [Fortsch.\ Phys.\  {\bf 29} (1981) 505];\\
  M.~A.~Shifman and M.~I.~Vysotsky,
  Nucl.\ Phys.\  B {\bf 186} (1981) 475;\\
  A.~P.~Bukhvostov, G.~V.~Frolov, L.~N.~Lipatov and E.~A.~Kuraev,
  Nucl.\ Phys.\  B {\bf 258} (1985) 601.
%
\bibitem{LO1}
  J.~Bl\"umlein,
  Eur.\ Phys.\ J.\  C {\bf 20} (2001) 683
  [arXiv:hep-ph/0104099].
%
\bibitem{LO2}
  A.~Mukherjee and D.~Chakrabarti,
  Phys.\ Lett.\  B {\bf 506} (2001) 283
  [arXiv:hep-ph/0102003].
%
\bibitem{LXLO}
  R.~Kirschner, L.~Mankiewicz, A.~Schafer and L.~Szymanowski,
  Z.\ Phys.\  C {\bf 74} (1997) 501
  [arXiv:hep-ph/9606267].
%
\bibitem{NLO}
  A.~Hayashigaki, Y.~Kanazawa and Y.~Koike,
  Phys.\ Rev.\  D {\bf 56} (1997) 7350
  [arXiv:hep-ph/9707208].
%
\bibitem{NLO2}
  S.~Kumano and M.~Miyama,
  Phys.\ Rev.\  D {\bf 56} (1997) 2504
  [arXiv:hep-ph/9706420].
%
\bibitem{NLO1}
  W.~Vogelsang,
  Phys.\ Rev.\  D {\bf 57} (1998) 1886
  [arXiv:hep-ph/9706511] and references therein.
%
\bibitem{NFW}
  A.~V.~Belitsky and D.~M\"uller,
  Phys.\ Lett.\  B {\bf 417} (1998) 129
  [arXiv:hep-ph/9709379];\\
  P.~Hoodbhoy and X.~D.~Ji,
  Phys.\ Rev.\  D {\bf 58}, 054006 (1998)
  [arXiv:hep-ph/9801369];\\
  A.~V.~Belitsky, A.~Freund and D.~M\"uller,
  Phys.\ Lett.\  B {\bf 493} (2000) 341
  [arXiv:hep-ph/0008005].
%
\bibitem{IK}
  B.~L.~Ioffe and A.~Khodjamirian,
  Phys.\ Rev.\  D {\bf 51} (1995) 3373
  [arXiv:hep-ph/9403371].
%
\bibitem{GRAC}
  J.~A.~Gracey,
  Nucl.\ Phys.\  B {\bf 662} (2003) 247
  [arXiv:hep-ph/0304113];
  Nucl.\ Phys.\  B {\bf 667} (2003) 242
  [arXiv:hep-ph/0306163];
  JHEP {\bf 0610} (2006) 040
  [arXiv:hep-ph/0609231];
  Phys.\ Lett.\  B {\bf 643} (2006) 374
  [arXiv:hep-ph/0611071].
%
\bibitem{VW}
  W.~Vogelsang and A.~Weber,
  Phys.\ Rev.\  D {\bf 48} (1993) 2073.
%
\bibitem{soft}
  H.~Shimizu, G.~Sterman, W.~Vogelsang and H.~Yokoya,
  Phys.\ Rev.\  D {\bf 71} (2005) 114007
  [arXiv:hep-ph/0503270].
%
\bibitem{BUZA1}
  M.~Buza, Y.~Matiounine, J.~Smith, R.~Migneron and W.~L.~van Neerven,
  Nucl.\ Phys.\  B {\bf 472} (1996) 611
  [arXiv:hep-ph/9601302].
%
\bibitem{AHF2}
  M.~Buza, Y.~Matiounine, J.~Smith and W.~L.~van Neerven,
  Nucl.\ Phys.\  B {\bf 485} (1997) 420
  [arXiv:hep-ph/9608342];\\
  M.~Buza, Y.~Matiounine, J.~Smith and W.~L.~van Neerven,
  Eur.\ Phys.\ J.\  C {\bf 1} (1998) 301
  [arXiv:hep-ph/9612398];\\
  I.~Bierenbaum, J.~Bl\"umlein and S.~Klein,
  Phys.\ Lett.\  B {\bf 672} (2009) 401
  [arXiv:0901.0669 [hep-ph]];
  Phys.\ Lett.\  B {\bf 648} (2007) 195
  [arXiv:hep-ph/0702265];
  Acta Phys.\ Polon.\  B {\bf 38} (2007) 3543
  [arXiv:0710.3348 [hep-ph]];\\
  I.~Bierenbaum, J.~Bl\"umlein, S.~Klein and C.~Schneider,
  Nucl.\ Phys.\  B {\bf 803} (2008) 1
  [arXiv:0803.0273 [hep-ph]].
%
\bibitem{BBK1}
I.~Bierenbaum, J.~Bl\"umlein and S.~Klein,
  Nucl.\ Phys.\  B {\bf 780} (2007) 40
  [arXiv:hep-ph/0703285].
%
\bibitem{BFNK}
  J.~Bl\"umlein, A.~De Freitas, W.~L.~van Neerven and S.~Klein,
  Nucl.\ Phys.\  B {\bf 755} (2006) 272
  [arXiv:hep-ph/0608024].
%
\bibitem{BBK3}
  I.~Bierenbaum, J.~Bl\"umlein and S.~Klein,
  Nucl.\ Phys.\  B {\bf 820} (2009) 417
  [arXiv:0904.3563 [hep-ph]].
%
\bibitem{AHF3}
  I.~Bierenbaum, J.~Bl\"umlein and S.~Klein,
  PoS {\sf CONFINEMENT8} (2008) 185
  [arXiv:0812.2427 [hep-ph]];
  Nucl.\ Phys.\ Proc.\ Suppl.\  {\bf 183} (2008) 162
  [arXiv:0806.4613 [hep-ph]].
%
\bibitem{SOFFER}
  J.~Soffer,
  Phys.\ Rev.\ Lett.\  {\bf 74} (1995) 1292
  [arXiv:hep-ph/9409254];\\
  D.~W.~Sivers,
  Phys.\ Rev.\  D {\bf 51} (1995) 4880;\\
  G.~R.~Goldstein, R.~L.~Jaffe and X.~D.~Ji,
  Phys.\ Rev.\  D {\bf 52} (1995) 5006
  [arXiv:hep-ph/9501297].
%
\bibitem{JI}
  X.~D.~Ji,
  Phys.\ Rev.\  D {\bf 49} (1994) 114
  [arXiv:hep-ph/9307235].
%
\bibitem{BAMU}
  A.~Bacchetta and P.~J.~Mulders,
  Phys.\ Rev.\  D {\bf 62} (2000) 114004
  [arXiv:hep-ph/0007120].
%
\bibitem{EIC}
C. Aidala et al. {\sf A High Luminosity, High Energy Electron-Ion-Collider}, 
A White Paper Prepared for the NSAC LRP 2007.
%
\bibitem{HSUM}
  J.~Bl\"umlein and S.~Kurth,
  Phys.\ Rev.\  D {\bf 60} (1999) 014018
  [arXiv:hep-ph/9810241];\\
  J.~A.~M.~Vermaseren,
  Int.\ J.\ Mod.\ Phys.\  A {\bf 14} (1999) 2037
  [arXiv:hep-ph/9806280].
%
\bibitem{GWP}
  D.~J.~Gross and F.~Wilczek,
  Phys.\ Rev.\ Lett.\  {\bf 30} (1973) 1343;\\
  H.~D.~Politzer,
  Phys.\ Rev.\ Lett.\  {\bf 30} (1973) 1346.
%
\bibitem{TNW}
  R.~Tarrach,
  Nucl.\ Phys.\  B {\bf 183} (1981) 384;\\
  O.~Nachtmann and W.~Wetzel,
  Nucl.\ Phys.\  B {\bf 187} (1981) 333.
%
\bibitem{QGRAF}
  P.~Nogueira,
  J.\ Comput.\ Phys.\  {\bf 105} (1993) 279.
%
\bibitem{COLOR}
  T.~van Ritbergen, A.~N.~Schellekens and J.~A.~M.~Vermaseren,
  Int.\ J.\ Mod.\ Phys.\  A {\bf 14}  (1999) 41
  [arXiv:hep-ph/9802376].
%
\bibitem{FORM}
  J.~A.~M.~Vermaseren,
  arXiv:math-ph/0010025.
%
\bibitem{MATAD}
  M.~Steinhauser,
  Comput.\ Phys.\ Commun.\  {\bf 134} (2001) 335
  [arXiv:hep-ph/0009029].
%
\bibitem{BBG}
  J.~Bl\"umlein, H.~B\"ottcher and A.~Guffanti,
  Nucl.\ Phys.\  B {\bf 774} (2007) 182
  [arXiv:hep-ph/0607200];
  Nucl.\ Phys.\ Proc.\ Suppl.\  {\bf 135} (2004) 152
  [arXiv:hep-ph/0407089].
%
\bibitem{B12}
  W.~E.~Caswell,
  Phys.\ Rev.\ Lett.\  {\bf 33} (1974) 244;\\
  O.~V.~Tarasov, A.~A.~Vladimirov and A.~Y.~Zharkov,
  Phys.\ Lett.\  B {\bf 93} (1980) 429.
%
\bibitem{GAMV}
  S.~A.~Larin, T.~van Ritbergen and J.~A.~M.~Vermaseren,
  Nucl.\ Phys.\  B {\bf 427} (1994) 41;\\
  S.~A.~Larin, P.~Nogueira, T.~van Ritbergen and J.~A.~M.~Vermaseren,
  Nucl.\ Phys.\  B {\bf 492} (1997) 338
  [arXiv:hep-ph/9605317];\\
  A.~Retey and J.~A.~M.~Vermaseren,
  Nucl.\ Phys.\  B {\bf 604} (2001) 281
  [arXiv:hep-ph/0007294];\\
  J.~Bl\"umlein and J.~A.~M.~Vermaseren,
  Phys.\ Lett.\  B {\bf 606} (2005) 130
  [arXiv:hep-ph/0411111].
\end{thebibliography}
\end{document}